\documentclass[fleqn,usenatbib]{mnras}
\usepackage{esvect}

\usepackage{newtxtext,newtxmath}

\usepackage[T1]{fontenc}

\DeclareRobustCommand{\VAN}[3]{#2}
\let\VANthebibliography\thebibliography
\def\thebibliography{\DeclareRobustCommand{\VAN}[3]{##3}\VANthebibliography}

\usepackage{graphicx}	
\usepackage{amsmath}	
\usepackage[dvipsnames]{xcolor}

\title[LoTSS DR2 morphologies]{Radio-loud AGN morphology and host-galaxy properties in the LOFAR Two-Metre Sky Survey Data Release 2}

\author[L. Clews et al.]{
L. Clews$^{1}$\thanks{E-mail: lucy.clews@open.ac.uk},
J. H. Croston$^{1}$,
H. Dickinson$^{1}$,
B. Mingo$^{2}$,
M. J. Hardcastle$^{2}$,
B. Barkus$^{2}$,
\newauthor
J. M. G. H. J. de Jong$^{3}$
and H. J. A. R\"ottgering$^{3}$
\\
$^{1}$School of Physical Sciences, Open University, Walton Hall, MK7 6AA, UK.\\ $^{2}$Centre for Astrophysics Research, University of Hertfordshire, College Lane, Hatfield AL10 9AB, UK.\\
$^3$ Leiden Observatory, Leiden University, PO Box 9513, 2300 RA Leiden, The Netherlands.
}

\date{Accepted XXX. Received YYY; in original form ZZZ}

\pubyear{\the\year{}}

\begin{document}
\label{firstpage}
\pagerange{\pageref{firstpage}--\pageref{lastpage}}
\maketitle

\begin{abstract}

Radio-loud active galaxies (RLAGN) can exhibit various morphologies. The Fanaroff-Riley (FR) classifications, which are defined by the locations of peaks in surface brightness, have been applied to many catalogues of RLAGN. The FR classifications were initially found to correlate with radio luminosity. However, recent surveys have demonstrated that radio luminosity alone does not reliably predict radio morphology. We have devised a new-semi automated method involving ridgeline characterisations to compile the largest known classified catalogue of RLAGN to date with data from the second data release of the LOFAR Two-Metre Sky Survey (LoTSS DR2). We reassess the FR divide and its cause by examining the physical and host galaxy properties of $3590$ FRIIs and $2354$ FRIs (at $z \leq 0.8$). We find that RLAGN near the FR divide with $10^{25} \le L_{144} \le 10^{26} $ WHz$^{-1}$ are more likely to show FRI over FRII morphology if they occupy more massive host galaxies. We find no correlation, when considering selection effects, between the FR break  luminosity and stellar mass or host-galaxy rest-frame absolute magnitude. Overall, we find the cause of different radio morphologies in this sample to be complex. Considering sources near the FR divide with $10^{25} \le L_{144} \le 10^{26} $ WHz$^{-1}$, we find evidence to support inner environment having a role in determining jet disruption. We make available a public catalogue of morphologies for our sample, which will be of use for future investigations of RLAGN and their impact on their surroundings.

\end{abstract}




\section{Introduction}\label{introduction}
Active galactic nuclei (AGN) are compact regions at the centre of some galaxies that emit radiation across the electromagnetic spectrum. AGN that have large regions of radio emission (also known as radio lobes) that can extend well beyond the visible structure of the host galaxy are known as radio-loud AGN (RLAGN) or radio galaxies. These regions are driven by collimated outflows of plasma from the AGN, known as radio jets, at relativistic speeds on pc to Mpc scales  \citep[see e.g.][for an in-depth review]{Bridle1984, 1984Begelman, Hardcastle2020}. 

Extended RLAGN can exhibit various morphologies \citep[e.g.][]{1980Miley}, but will most commonly consist of one of two categories established by \cite{F+R1974}: Fanaroff-Riley (FR) type 1 (FRI) or type 2 (FRII). FRI sources are defined to be centre-brightened, whilst FRII sources are edge-brightened. The FR classifications are widely used and have been applied to numerous surveys. With these classifications, \cite{F+R1974} suggested a correlation between the relative positions of high and low radio surface brightness (flux density per unit area) regions of RLAGN and radio luminosity in the $3$CR sample \citep{1971Mackay}.

The physical cause of the FR divide remains under debate. For example, it has been suggested that accretion mode could influence jet morphology. A direct link between the two would mean that the properties of the launched jet are determined by the AGN accretion disc or flow, such that jets are intrinsically different when they are launched. This is also known as the central engine model \citep{1997Jackson, 2011Meyer, 2021Keenan}. However, this model cannot explain why jets of the same inferred power are able to turn into different morphologies \citep{Hardcastle2020, Mingo2022}. FRIIs are thought to remain relativistic for the full length of their jets until terminating in a hotspot, while FRIs are known to be initially relativistic and then become mass-loaded due to entrainment, causing them to decelerate on kpc scales \citep{Bicknell1994, 2002Laing, 2015Wykes,Tchekhovskoy2016, Hardcastle2020}.  Properties of the central engine alone cannot account for this structural difference. Instead, observations of kpc-scale jet deceleration suggest a relationship between jet power and host-scale environmental density such that jets of the same initial power will remain collimated and relativistic in a poor environment and decelerate, entrain material and expand in a richer one. 

There is some observational evidence to support this theory. \cite{Ledlow+Owen1996} found the FRI/II luminosity break to be dependent on host-galaxy magnitude, such that FRIs have higher radio luminosities when in brighter host galaxies (where the density of the interstellar medium is assumed to be
higher). However, this result has since been questioned  due to severe selection effects in the sample \citep{Best2009, Lin_2010, Capetti2017, Hardcastle2020}, so it is not known whether it holds for the full population of RLAGN. In addition, interpreting the results of \cite{Ledlow+Owen1996} in the context of a jet disruption model relies on radio luminosity being a good proxy for jet power. Differences in particle content, external pressure and radiative losses mean that this conversion is known to have large systematic biases and scatter (for more details see section 4.5 of \citeauthor{Hardcastle2020} \citeyear{Hardcastle2020} or \citeauthor{croston2018} \citeyear{croston2018}, \citeauthor{2018Hardcastlesim} \citeyear{2018Hardcastlesim}).

The result also uses host-galaxy absolute magnitude as an estimator for the density of material available to disrupt jets. Mass loading in FRIs can occur by different methods. The primary mechanism considered in the literature is external entrainment from hot-gas in the surrounding interstellar medium \citep[ISM,][]{Bicknell1994, 2002Laing}. For massive elliptical galaxies, which are the typical hosts of RLAGN, the host-galaxy environmental richness is expected to scale with galaxy mass \citep{2005Best, 2013Kim}, for which absolute magnitude is a proxy \citep{Kondapally2022}. An
alternative mechanism is internal entrainment, where mass is loaded
from stellar winds from stars embedded within the jet. In this scenario, the density of young
stellar populations could be more important than the host-galaxy
stellar mass. Mass loading from stellar winds has been argued to be sufficient to decelerate FRI-jets in some cases \citep{1994Komissarov,1996Bowman,2006Hubbard, 2015Wykes}. However, \cite{2014Perucho} found for FRIs of typical power ($L_{j} \approx 10^{43}-10^{44}$ erg s$^{-1}$),  stellar winds alone do not produce
significant deceleration. 

 In more recent years as radio surveys have improved to reach lower flux limits, evidence has emerged that morphology is not as closely related to luminosity as originally believed. An unexpected population of low-luminosity FRIIs (FRII-lows) was recently discovered \citep{Best2009, Capetti2017, Mingo2019, Mingo2022}, which form a substantial proportion of FRIIs below $z \sim 0.8$ \citep{Mingo2019, Mingo2022}. It is now known that there is a large overlap in the luminosity distributions of FRIs and FRIIs \citep{2010Gendre, Mingo2019, Mingo2022, 2024Jurjen}, complicating the interpretation of the two different morphologies and how they are caused. Currently, evidence points to FRII-lows being a mix of RLAGN populations \citep{Mingo2019, Mingo2022}. One theory is that FRII-lows could occupy lower-mass hosts, so can remain collimated even though they have a relatively low power \citep{Mingo2019}. Another possibility is that FRII-lows are no longer be active, just fading. In this case, they would not be comparable to active sources. Comparisons have been made between FRII-lows and FRIs of similar luminosities and physical size. The probability of a jet of intermediate luminosity becoming an FRI or FRII was shown to depend on the stellar mass of the host galaxy \citep{Mingo2022}. A relationship between jet power and host-galaxy magnitude has not yet been ruled out; \cite{Mingo2019} showed some evidence to support the Ledlow-Owen relation, but could not rule out that the observed relation was due to selection effects.

Host-galaxy scale environmental interactions with the jet can impact source morphology. For some low power sources, large-scale environmental interactions can cause the outer portion of the jet to bend or distort away from a linear trajectory. This subclass of FRIs are known as bent-tailed (BT) sources \citep{OBrien2018, Mingo2019}, which can be subclassified based on their opening angle, $\theta_{\rm open}$. BTs with $\theta_{\rm open} > 90^\circ$ are known as wide-angle tailed (WAT) sources \citep{OwenRudnick1976, Mingo2019} whilst BTs with $\theta_{\rm open} < 90^\circ$ are narrow-angle tailed (NAT) \citep{RudnickOwen1976, 1985ODea,Mingo2019}. The favoured explanation for BTs is that they form in dense environments such as galaxy clusters \citep{Owen1976,Blanton2000,Blanton2001,Wing2011,Morris2022, DeGazon2023}, where they are bent by ram pressure \citep{Jones+Oweb1979,Burns1980,Blanton2000,Morsony2013}. As a result, BTs have been used to identify galaxy clusters at high redshift \citep{Wing2011, 2015Blanton, golden-marx2021}. However, BTs have also been found in galaxy groups \citep{Ekers1978,Freeland2011}, in filaments \citep{Edwards2010} and in isolated environments where the ICM is not likely to be dense enough to generate ram pressure \citep{Blanton2001}.

The relationship between morphology and environment is clearly vital to understanding the FR break and the diversity of the RLAGN population. The LOFAR Two-Metre Sky Survey (LoTSS) provides a unique opportunity to explore the RLAGN population using the largest sample to date, with unmatched sensitivity to extended structure on $6$ arcsec scales. Our motivation for this work is to derive an automated classification technique suited to large samples, which produces groups of sources with similar jet morphologies and low contamination, motivated by the physical properties of each source. A large morphologically classified catalogue has the potential to greatly improve our understanding of how radio sources evolve and affect their environment under different conditions at different redshifts. This could then be applied to exploring the locations and mechanisms of AGN feedback. As such, we want to explore all of the information available rather than applying a standardised set of rules that would produce binary classifications. The aim is to produce clean classifications of objects, rather than to obtain complete population statistics or classify every single source.

This paper is structured as follows: in Section \ref{section: Data} we describe the details of our dataset derived from LoTSS DR2 \citep{LoTSSDR22022}. In Section \ref{section: classification} we outline our classification method, and then present the morphological and environmental properties of our sample in Section \ref{results}. In Section \ref{discussion} we discuss our interpretation of the results for the different populations before presenting conclusions in Section \ref{conclusions}.

\section{Data}\label{section: Data}
This section outlines how we derived a sample of RLAGN suitable for automated classification. Section \ref{sec:AGN sample} describes our starting catalogue. Section \ref{ridgelines} explains our decision to use ridgelines \citep{Barkus2022} for this method and the selection criteria applied to them. We then explain how we use ridgelines to extract structural and morphological information about the sources in our sample in Section \ref{section: ridgeline analysis} and \ref{curvature method}.

\subsection{AGN sample selection}\label{sec:AGN sample}

The datasets used in this work are from the public second data release of the LOw Frequency ARray (LOFAR) Two Metre Sky Survey (LoTSS DR$2$: \citealt{LoTSSDR22022}). LoTSS is an ongoing survey of the northern sky at $120-168$ MHz with $6$ arcsec resolution. LoTSS DR$2$ contains over $4$ million sources across $27\%$ of the northern sky. Of these, $85\%$ have optical host galaxy identifications and $58\%$ have good spectroscopic or photometric redshifts \citep{Williams2019, 2023HardcastleLoTSShosts}. For further details and validation of the quality of host-galaxy identifications, see \cite{2023HardcastleLoTSShosts}.

We make use of the $6$ arcsec Stokes I images, which have a central frequency of $144$ MHz and median rms sensitivity of $83$ $\mu$Jy/beam. The calibration process of the LoTSS fields is described by \cite{LoTSSDR22022} and treats both direction-dependent and direction-independent effects \citep{2021Tasse}. We restrict our sample to the LoTSS DR2 AGN catalogue \citep{2025Hardcastle}, which contains $666,804$ sources. This catalogue (hereafter known as H$25$) was designed to select a clean sample of radio-loud AGN.

\subsection{Ridgeline data}\label{ridgelines}

 Ridgelines trace the pathway of highest radio flux density along a jet \citep[e.g.][]{Barkus2022}. Each point along the ridgeline is calculated as the maximum flux along an annular slice at a distance set by the user (in this case the beam size of LOFAR),  within the set search sector (of half-size $60^{\circ}$).
The maximum ridgeline length is set to $0.95$ $\times$ source size from the initial point, in both search directions. Preliminary work by \cite{Barkus2022} indicated the usefulness of ridgelines for morphological characterisation - they record surface brightness (SB) information and provide a unique measure of jet size as the distance over which material has travelled along the jet path. Ridgelines are expected to pass through the centre of the optical/infrared host galaxy of a radio source and have been applied as part the host galaxy identification process for LoTSS DR$2$ \citep{2023HardcastleLoTSShosts}.

 The ridgeline code, RL-Xid\footnotemark \footnotetext{\url{https://github.com/BonnyBlu/RL-Xid}}, has been applied to a subset of extended sources in LoTSS DR2 ($>15$ arcsec and $>10$ mJy), for which they were expected to provide useful morphological information (described by H$24$). In this work, we use ridgelines to find clean samples of RLAGN with similar jet morphologies. Our initial sample consists of all sources in H$24$ that have a ridgeline ($56,161$). These sources have between $3$ and $35$ points along their ridgeline, but most ($94.4\%$) have fewer than $20$.  Due to our method uniquely relying on ridgelines, it was vital that we apply some initial ridgeline quality filters. We chose to limit our sample to sources in H$24$ with $>5$ ridgeline points, to ensure each source was sufficiently well sampled for our analysis. After this filter there were $50,234$ sources remaining. We discuss the effect of angular size-related cuts on our sample selection and results in Section \ref{sec: selection effects}.
 



\subsection{Ridgeline analysis}\label{section: ridgeline analysis}

We fit a spline model with natural boundary conditions to each ridgeline in our sample to smooth them as well as to obtain a differentiable function that describes the ridgeline. Due to the complex nature of RLAGN, the ridgelines that trace them can have bends of varying size and degree. It can be impossible to set up a Cartesian coordinate system where all points on the ridgeline have unique values of both the $x$ and $y$ coordinate. Instead, the ridgelines are best characterised by a parametric spline. We defined a third order, uniform parametrisation with $200$ linearly spaced points corresponding to linearly spaced values of the spline parameter. We chose to use $200$ spline points as this is much greater than the number of points on any ridgeline (which will have maximum $\sim 30$ points) by a large enough margin that the spline would be interpolating across a very small distance with each iteration, resulting in a smooth representation of the ridgeline. The spline had to be differentiable to at least second order for later analysis (see Section \ref{curvature method}), so a polynomial of minimum order $3$ was required. We experimented with higher orders and found they did not produce smoother representations of the ridgelines.
 

We also fitted an interpolating cubic spline to the SB profiles generated by RL-Xid for each source in our sample of equal length to the parametric ridgeline splines. The SB of an individual pixel in RL-Xid is defined as the weighted average of the centre pixel and the $4$ orthogonally adjacent pixels. To exclude sources with low signal to noise that would be difficult to classify, we limited our sample to sources with a peak to mean SB ratio (dynamic range) greater than $2$, leaving $49,799$  sources with SB information. These SB spline profiles were used in the first stage of our classification method, described in Section \ref{section: classification}

\subsection{Curvature analysis}\label{curvature method}
As an additional means of classification, we chose to quantify the curvature ($\kappa$) of the ridgelines, to explore the properties of BTs. The curvature of a line can be measured over short increments (or arclengths, $s$) along it as a function, $f(x)$, of $\psi$, the angle which a tangent to the ridgeline at that point makes with the positive x axis. The definition of $\kappa$ used in this paper is,

\begin{equation}
    \kappa = \frac{d\psi}{ds} = \frac{d\psi}{dx}/ \frac{ds}{dx} = \frac{f''(x)}{(1+f'(x)^2)^{3/2}}
    \label{curvature roc}
\end{equation}

where 
\begin{equation}
    \nonumber
    \frac{d\psi}{dx}= \frac{f''(x)}{(1+f'(x)^2)}, \frac{ds}{dx}=(1+f'(x)^2)^{1/2}
\end{equation}

In our case the function describing the ridgeline is unknown. Instead we have a differentiable parametric spline model (of order $k=3$ and  $200$ points) for each ridgeline. We use the parametric curvature equation given by,

\begin{equation}\label{curvature equation}
    \kappa = \frac{\dot{x}\ddot{y}-\dot{y}\ddot{x}}{(\dot{x}^2+\dot{y}^2)^{3/2}}
\end{equation}

where $u$ is a uniform parameter, $\dot{x}=dx/du, \dot{y}=dy/du, \ddot{x}=d^2x/du^2$ and $\ddot{y}=d^2y/du^2$.

\begin{figure*}
    \centering
    \includegraphics[width=\linewidth]{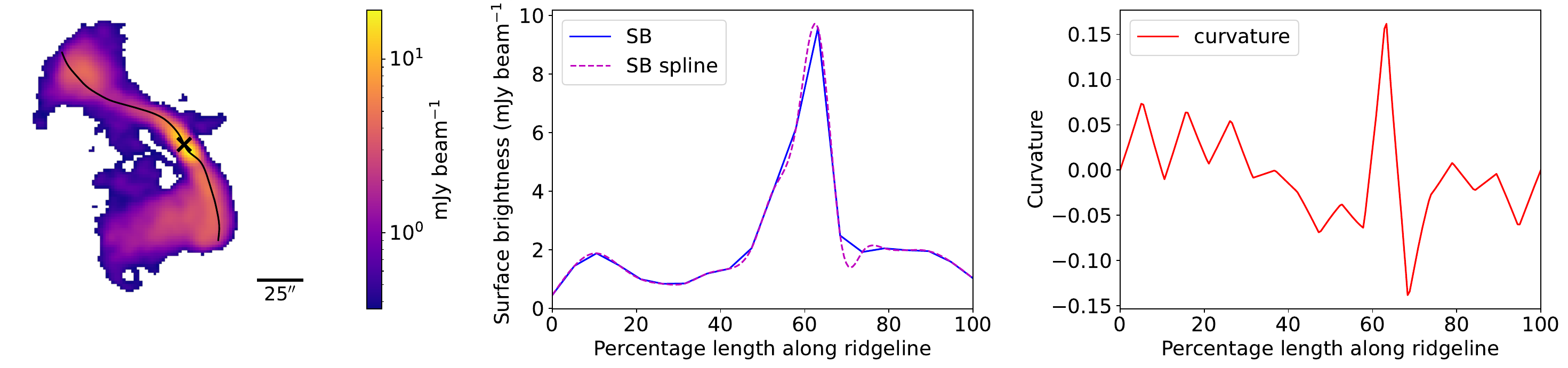}
    \caption{Example of input data. Left: LoTSS DR$2$ source ILTJ$090543.94+422105.9$ with ridgeline spline (black line) and closest ridgeline spline point to the host galaxy position (black cross) indicated. Middle: Surface brightness profile obtained from RL-Xid (solid blue) overlaid with its spline model (dashed magenta). Right: curvature profile (red). }
    \label{fig: input data}
\end{figure*}

For every source in our sample we obtain a curvature profile.  Fig. \ref{fig: input data} shows an example of the structural information (ridgeline spline, SB spline profile and curvature) extracted from the ridgelines for each source in our sample.
Both the magnitude and direction of curvature along the jet path are useful quantities to classify the morphology of RLAGN. We incorporate curvature as a secondary step in our classification process (described in in Section \ref{section: classification}).

\section{Classification method}\label{section: classification}
The aim of this work is to identify clean samples of objects with similar jet morphologies, whilst accounting for their physical properties, in an automated way. This section outlines our classification method, which uses the ridgeline structure information described in the previous section. In Section \ref{dimensionality reduction and clustering} we describe how we reduced the dimensionality of our surface brightness spline profiles before using unsupervised density-based clustering to place them into groups with similar features. Next, we explain how host galaxy locations were incorporated into classifying morphology in Section \ref{method: host subgroups} (with more detail in Appendix A). We then outline our method for calculating the jet opening angle on a subset of sources and its usefulness in identifying BTs in Section \ref{opening angles}. We then explain how final morphological classifications were assigned using a combination of surface brightness, host location, curvature and opening angles in Section \ref{final classification}. Lastly, in Section \ref{sec: selection effects}, we consider selection effects.

\subsection{Dimensionality reduction and clustering}\label{dimensionality reduction and clustering}
We first used the Uniform Manifold Approximation and Projection algorithm \citep[UMAP,][]{umap2018} to embed our $200$-dimensional SB spline data into two dimensions in a way that best preserves the overall structure of the higher-dimensional data. We used the \textsc{umap-learn} \footnotemark \footnotetext{\url{https://github.com/lmcinnes/umap}} implementation of the UMAP algorithm using the parameters listed in Table \ref{Overview of the UMAP and HDBSCAN parameters used.}.

The next step in our method was to separate the RLAGN in our sample into clusters with similar features in their SB spline profiles, without feeding in any preconceptions of known classes. Due to the nature of this approach, the number of clusters expected was not known. We chose to use the Hierarchical Density-Based Spatial Clustering of Applications with Noise (HDBSCAN) algorithm \citep{mcinnes2017hdbscan} as it does not require prior knowledge of how many clusters are expected and allows for groups of varying size and densities, unlike some other density-based clustering algorithms.

This technique has not been used to group LOFAR RLAGN before. Other automated methods have inputted LOFAR images directly into neural networks \citep[e.g.][]{2019Lukic, 2021Mostert, 2025Baron}. This method uses ridgelines to parametrise the images before using them for classification. The advantage of this method is that it should separate sources into groups with similar jet morphologies, which can then be used to study jet evolution under different conditions.

\begin{figure*}
    \includegraphics[width=0.33\linewidth]{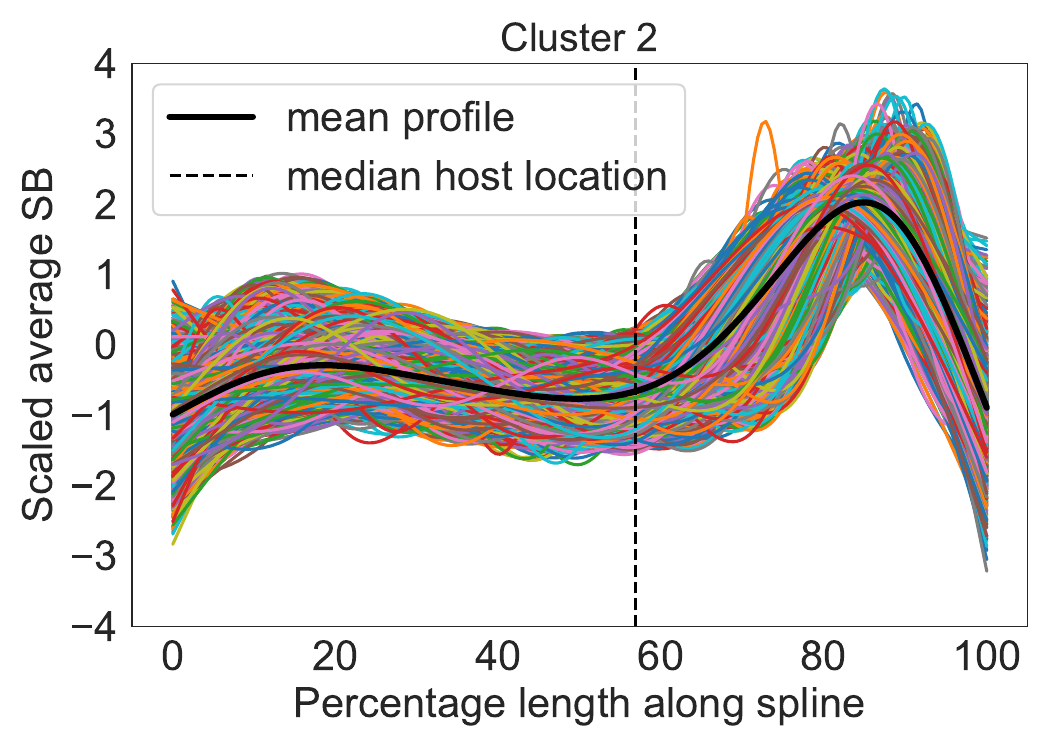}\raisebox{0.4cm}{\includegraphics[scale=0.55]{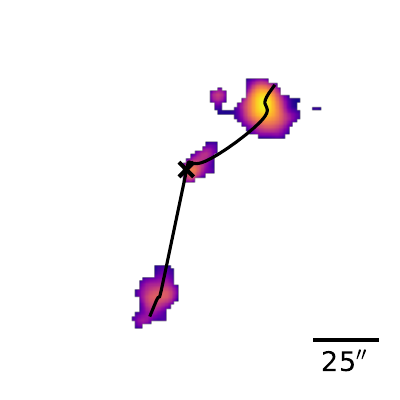} \includegraphics[scale=0.55]{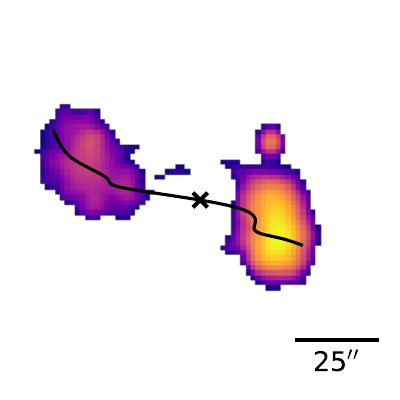}
    \includegraphics[scale=0.55]{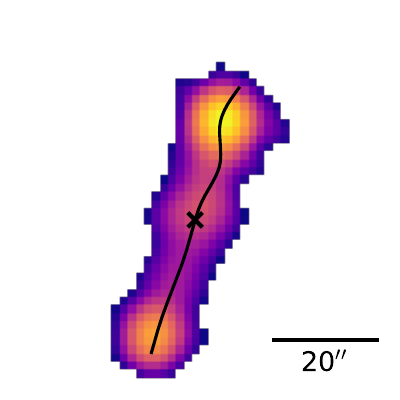}} \\
    \includegraphics[width=0.33\linewidth]
{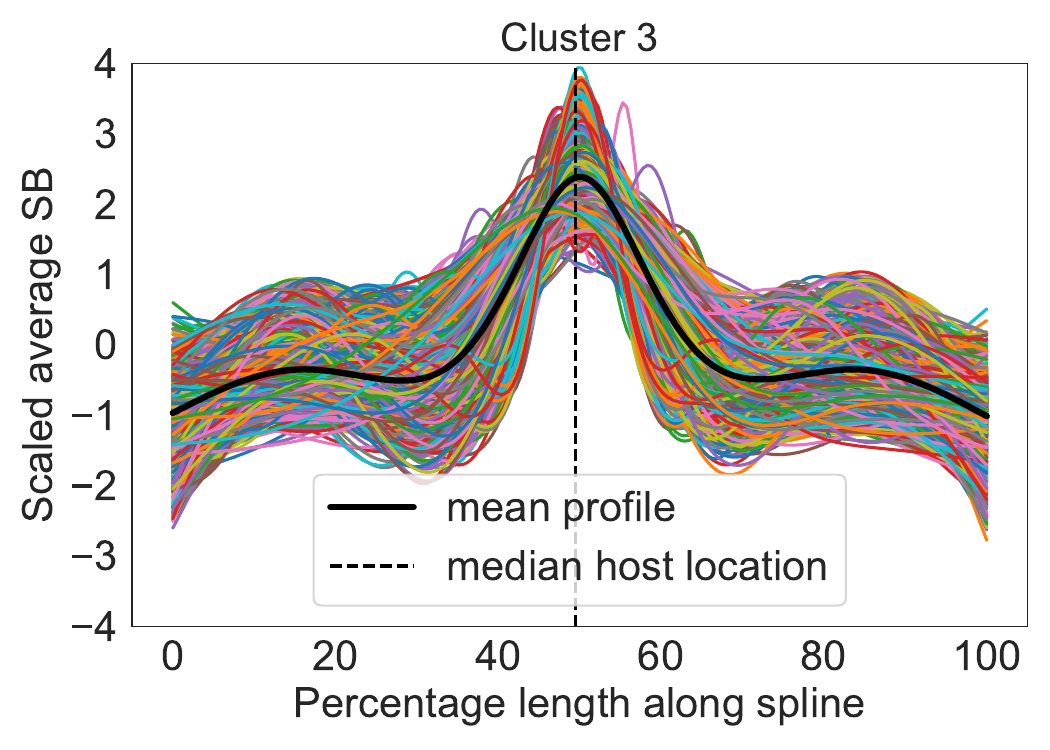} 
   \raisebox{0.4cm}{\includegraphics[scale=0.55]{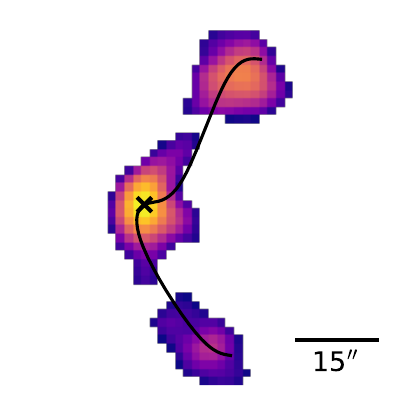} 
   \includegraphics[scale=0.55]{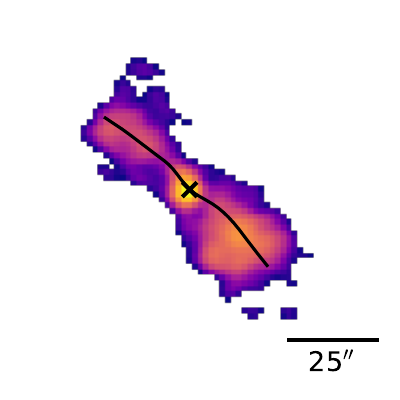}
    \includegraphics[scale=0.55]{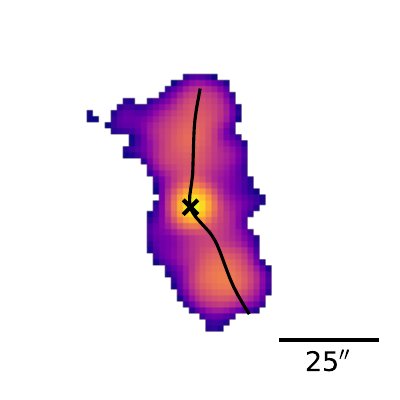}} \\
    
    \caption{Left: Two example surface brightness clusters generated by HDBSCAN, corresponding to FRII-like (top) and FRI-like (bottom) sources. Each surface brightness profile is scaled by z-score standardisation and represented by a different coloured line. The solid black line on each figure represents the mean surface brightness profile of that cluster. The vertical dashed line represents the median host galaxy location of the cluster. Right: Three randomly chosen example sources from the SB clusters, overlaid with their ridgeline spline (black line). The black cross indicates the location of the closest ridgeline spline point to the host galaxy position. }
    \label{figure: example SB groups}
\end{figure*}

We ran HDBSCAN on our previously defined UMAP embedding using the parameters in Table \ref{Overview of the UMAP and HDBSCAN parameters used.}. We experimented with the HDBSCAN parameters until we found values that produced clusters that contained sufficiently similar SB profile characteristics - i.e. numbers and locations of SB peaks, overall shape, and whether `nearby' clusters appeared sufficiently distinct. With these parameters, HDBSCAN clustered $82.1\%$ of our sample ($40,903$ sources) into $25$ clusters of varying size, the remaining sources were classified by HDBSCAN as `noise'. Fig. \ref{figure: example SB groups} shows two example SB cluster groups and images of three randomly selected sources from within them. We visually inspected $100$ sources from each cluster group to check if they were morphologically similar. We estimated that typically $70\%$ of sources in each cluster group were morphologically similar before host galaxy locations were considered. We also found that some of the HDBSCAN groups had very similar morphological characteristics to each other, e.g. with reflected locations of peak brightness. To reduce the amount of contamination we calculated the mean SB spline profile for each cluster group and reassigned sources with $5$ or more SB spline points that were $3$ or more standard deviations away from the mean profile of their original cluster, to the cluster group they fit best. The new best fit group was determined by the smallest value of the mean squared error (MSE) between the outlying profile and the mean spline profile of each SB cluster. We chose this threshold of spline points and number of standard deviations after experimenting with different values, each time checking that the remaining profiles had similar shapes and obvious large outliers were identified.

After this reassignment, we again visually inspected a random sample of $100$ sources from each SB cluster. We found that the remaining contamination within the HDBSCAN cluster groups was from sources of smaller angular size, or sources with an unusual host galaxy location. As the aim of this work is to produce clean classes of sources, we applied an additional filter to restrict our final sample to sources with largest angular size (LAS) $>45$ arcsec. We discuss the potential implications of this on our scientific conclusions and future work in Section \ref{sec: selection effects}.

\subsection{Host subgroups}\label{method: host subgroups}

There may be classes of objects that have similar SB profiles but have host galaxies that are located at substantially different locations along them. The location of the host galaxy provides important information on the formation history and symmetry of the jets (the distance of each jet from the host galaxy). Sources with a host in the centre are a different class of objects to those with a host towards the end of the ridgeline. It is also possible that sources with hosts at uncommon locations compared to their SB class have incorrect host IDs. The next step in our classification pipeline was to incorporate the location of host galaxies to distinguish between situations with differing jet physics, and to identify outliers with incorrect IDs. 

We first calculated the closest point along the ridge spline to the host galaxy position and expressed this as a relative length along the ridgeline spline, to make host galaxy positions comparable among RLAGN with very different projected sizes on the sky. We then visually inspected the distribution of host galaxy locations along the ridgelines for each SB cluster identified by HDBSCAN.  A very small proportion of sources ($3\%$ of the total amount clustered) had host galaxies positioned within the first or last $10\%$ of their ridgeline splines. We visually inspected all of these sources and found the majority to have incomplete ridgelines or incorrect host IDs. We therefore decided to remove this small proportion of sources from our sample. We do not believe that this decision will discard BT sources that have been correctly traced by the ridgelines and therefore could have been used for our analysis. The ridgelines are designed to stop tracing when they either: find no additional points above a brightness threshold of four times the locally measured rms or reach the maximum ridgeline length \citep[defined by][]{Barkus2022}. In addition, we did not find correctly traced BTs in this group when performing the visual inspection.

We also found that for each cluster there were between $1$ and $3$ peaks in the distributions of host galaxy locations. We used  Gaussian mixture modelling (GMM) to separate these (see Appendix A). Sources with host galaxy locations that were within $1\sigma, 2\sigma$ and $3\sigma$ of their Gaussian group mean were assigned quality flags of Q$1$, Q$2$ and Q$3$ respectively. Sources whose host location was more than $3\sigma$ deviant were excluded. In total we had $20,172$ sources placed into host subgroups with a quality flag of Q$1-3$. We consider Q$1$ sources as `good quality' and only use these for scientific analysis in Section \ref{results}. Q$2$ and Q$3$ sources are not used in our analysis due to having atypical host locations in comparison to the rest of their surface brightness cluster, which could indicate that they have different jet dynamics. Nevertheless, we consider the impact of including Q$2$ sources on our conclusions in Section \ref{results}.

We again visually inspected a random sample of $100$ source images from each subgroup. Of each Q$1$ subgroup we typically found $ >90\%$ of sources to be morphologically similar. Of the sources that were not, most often the ridgeline did not represent the source accurately or appeared to have an incorrect host ID. We found that subgroups of Q$3$ were of lower purity, typically with $70\%$ or less of sources being morphologically similar. We explain how we use the host subgroups for classification in Section \ref{final classification}.

\subsection{Opening angles}\label{opening angles}
 We noticed some bent-tail (BT) sources within the FRI SB clusters when performing our visual inspections. This is to be expected because they have FRI-like gradually decreasing SB profiles.
BT sources are traditionally defined by their opening angle, $\theta_{\rm open}$. We used the ridgelines to explore $\theta_{\rm open}$ of the FRI SB clusters, to identify the BTs within them. We limited the sample for which $\theta_{\rm open}$ was to be calculated to the largest ($8$ or more ridgeline points) and most intrinsically curved sources (maximum absolute curvature amplitude $|\kappa_{\rm max}|>0.2$, see Section \ref{curvature method}) in the FRI subgroups found by our clustering method.

\begin{figure*}
    \centering
    \includegraphics[width=0.7\linewidth]{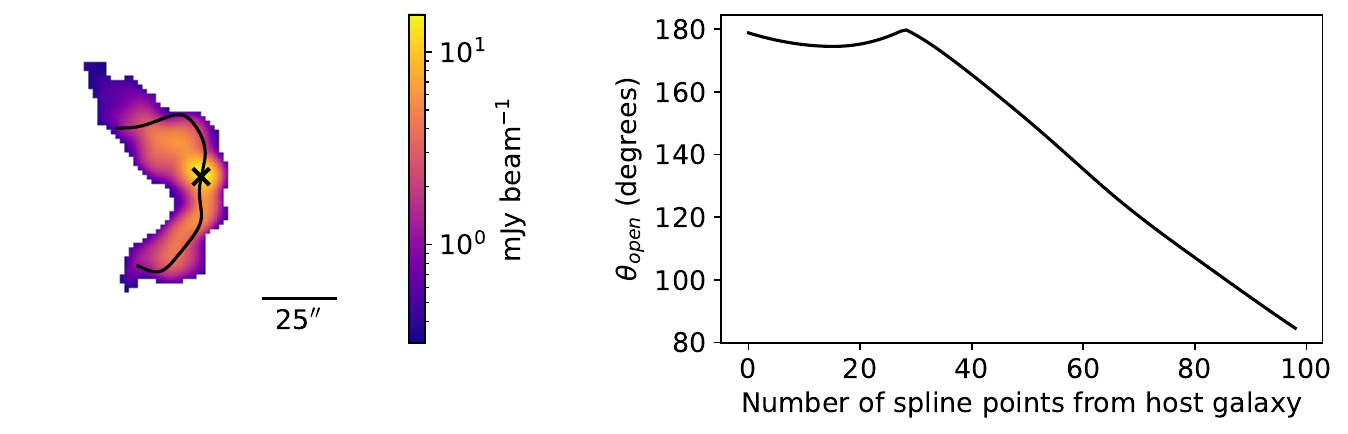}\\
    \caption{Left: Example bent source with ridgeline spline (black line) and closest ridgeline spline point to the host galaxy (black cross) shown. Right: Opening angle profile as a function of point number away from the host galaxy. }
    \label{fig: opening_angle profile examples}
\end{figure*}

The tails of BT sources are bent in the same direction. To filter for this, we used curvature information. Where the ridgeline splines were curved in the same direction (had opposite curvature magnitudes) at the first and last $10\%$, $\theta_{\rm open}$ was calculated by taking the dot product between two vectors $\vec{HA}$ and $\vec{HB}$:
\begin{equation}
   \theta_{\rm open} =  \cos^{-1}\left(\frac{\vec{HA} \cdot \vec{HB}}{|\vec{HA}||\vec{HB}|}\right)
    \label{opening angle formula}
\end{equation}
$\vec{HA}$ was initially calculated between the host galaxy (H) and the ridgeline spline point (A) immediately prior to H. $\vec{HB}$ was initially taken between H and the ridgeline spline point (B) immediately after H. The ridgeline spline points are spaced equally. With each iteration B and A were shifted by one step in the forward and backward direction along the ridgeline respectively. $\theta_{\rm open}$ was calculated during each iteration, until one end of the ridgeline was reached. We obtained an opening angle profile for each BT source candidate, beginning at the host galaxy, moving out towards the ends of the source. If the opening angle profile was found to decrease, the source was classified as `bent'. An example opening angle profile is shown in Fig. \ref{fig: opening_angle profile examples}.

We visually inspected all of the `bent' sources as there were only $223$ of them. We found all of them to have bent ridgelines, but estimate only $75\%$ of them to be bent sources. Where the source was not bent, the ridgeline did not follow the lobe or tail structure of the jet in the best way.  The focus in \cite{Barkus2022} was on the validity of the ridgelines for host identification, so they are not optimised for all classes of source. We describe how we use this information in the final stage of our classification method and planned future work in Section \ref{final classification}.

\subsection{Classification}\label{final classification}
\begin{table}
    \centering
    \begin{tabular}{l|c|l|c}
        \hline
        \multicolumn{2}{|c|}{UMAP parameters} & \multicolumn{2}{|c|}{HDBSCAN parameters}\\ 
        \hline
        Parameters & Number & Parameter & Number\\
        \hline
         number of neighbours & $400$ & minimum sample & $100$ \\
         
         minimum distance & $1\times10^{-3}$ & minimum cluster size & $800$ \\
         
         number of components & $2$ & maximum cluster size & $8500$ \\
         
         repulsion strength & $24$ \\
         \hline 
         
    \end{tabular}
    \caption{Overview of the UMAP and HDBSCAN parameters used.}
    \label{Overview of the UMAP and HDBSCAN parameters used.}
\end{table}

\begin{figure*}
    \centering
    \includegraphics[scale=0.52]{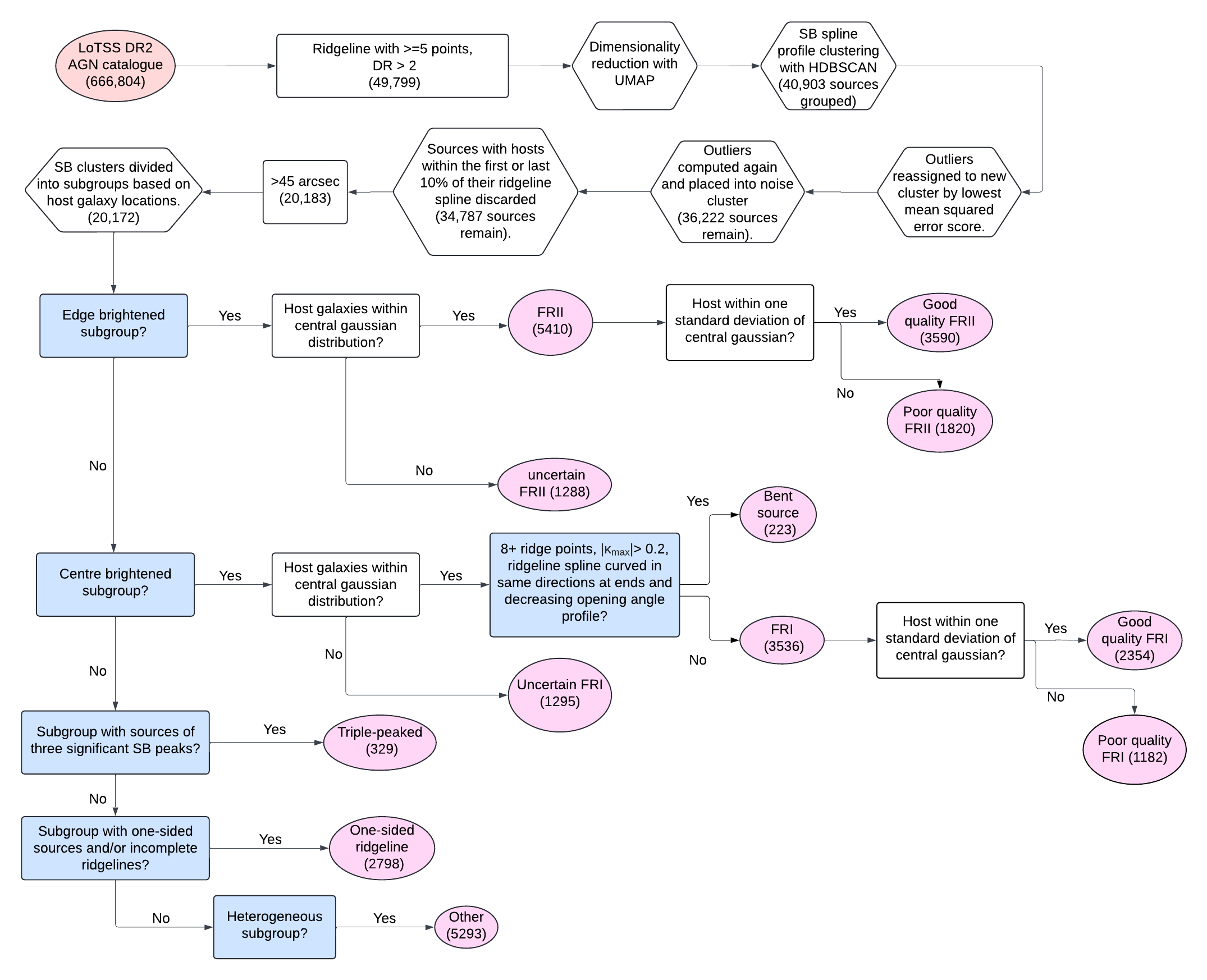}
    \caption{Flow chart outlining each stage in the classification process. Inputs and outputs are represented by orange and pink circles respectively. Blue rectangles indicate decisions about the final morphological labels. }
    \label{fig:classification flow chart}
\end{figure*}

At this stage in our method we had identified SB clusters using HDBSCAN, subgrouped these based on host galaxy positions and used ridgeline opening angles to identify `bent' sources within the FRI-like subgroups. This section describes how we combined all of the available information and assigned labels to sources. A flowchart of this process is shown in Fig. \ref{fig:classification flow chart} and final descriptions of classes are outlined in Table \ref{Table: description of final classes}. 

Edge-brightened subgroups where the host galaxy position was within the central Gaussian of its host distribution were labelled as `FRIIs', consisting of $5410$ sources. Of these FRIIs, $3590$ are labelled as Q$1$ FRIIs. These are sources where we can be most confident of the host ID. Edge-brightened subgroups where the host galaxy was not in the central Gaussian distribution were classified as `uncertain FRIIs'. This class has $1288$ sources.

We took all of the subgroups of centre-brightened morphology with host galaxies located in the central Gaussian of their distribution and combined them as having FRI-like morphology. We then used the method described in Section. \ref{opening angles} to identify $223$ `bent' sources. Of this `bent' class we estimate $75\%$ of sources to be intrinsically bent, as described in Section \ref{opening angles}. Despite this, we include them as a class to demonstrate the usefulness of the method which could be applied to BTs, S-shaped sources and to explore transitions between lobed and tailed sources. Future work is planned to improve RL-Xid to trace tailed emission more accurately and to filter out incorrect ridgelines to make this method more useful for selecting clean, science-ready source samples.

The remaining $3536$ sources in the FRI-like  morphology group not classified as `bent' were classified as FRIs, $2354$ of which are Q$1$ FRIs. Centre-brightened subgroups where the host galaxy was not in the central Gaussian distribution were classified as `uncertain FRIs', this class has $1295$ sources.

If the subgroup contained sources with $3$ SB peaks they were classified as `triple-peaked' sources. There then remained a number of subgroups where the sources were either themselves one-sided or were one-sided due to the ridgeline tracing only part of the source. By one-sided we mean that the source shows one bright peak and a tail of fainter emission. These sources were all placed into the `one-sided ridgeline' class, which contains $2798$ sources. From visually inspecting $100$ random sources from this class, we estimate that $\sim 60\%$ are one-sided sources, the rest being incomplete or incorrect ridgelines. To test whether the more uncommon source classes (triple-peaked and one-sided ridgelines) could be preferentially affected by redshift selection effects we compared their redshift distribution to the full sample, and did not find them to be more prevalent at high redshift. In addition, we visually inspected the host IDs of $100$ randomly chosen sources from these two classes and found only $5\%$ to be incorrect, consistent with the quality of the identifications for the full population \citep{2023HardcastleLoTSShosts}. Future work is planned to filter out incorrect ridgelines to make this class more useful for scientific investigation.

\begin{figure*}
    
    FRI \\
    \includegraphics[width=0.23\linewidth]{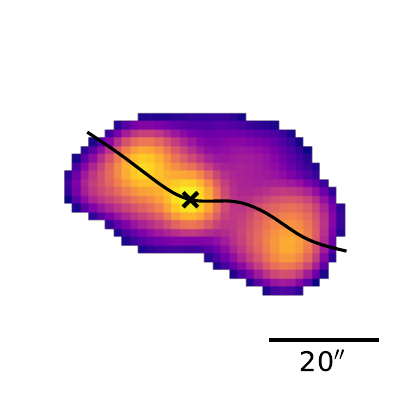}
    \includegraphics[width=0.23\linewidth]{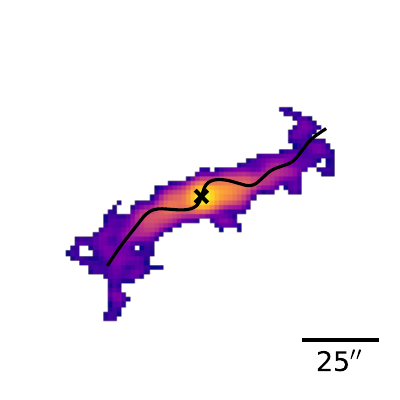}
    \includegraphics[width=0.23\linewidth]{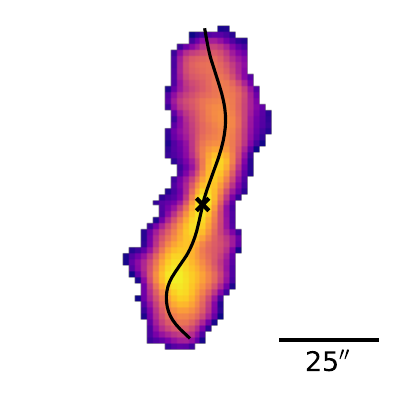}
    \includegraphics[width=0.23\linewidth]{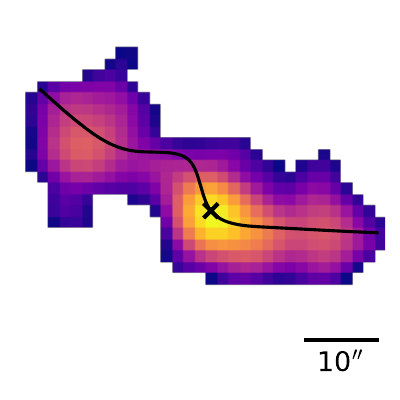} \\
    
    `bent'\\
    \includegraphics[width=0.23\linewidth]{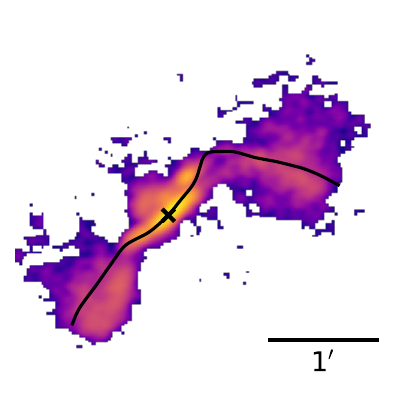}
    \includegraphics[width=0.23\linewidth]{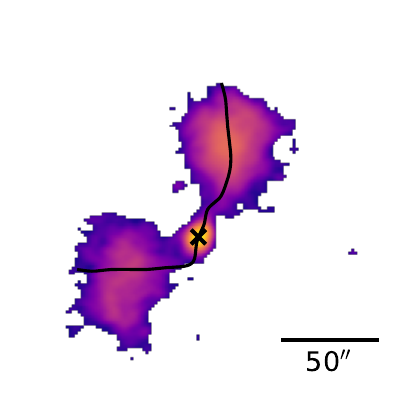}
    \includegraphics[width=0.23\linewidth]{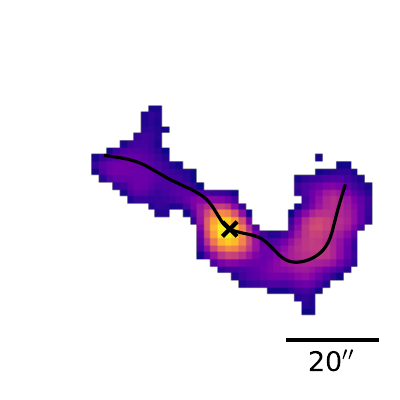}
    \includegraphics[width=0.23\linewidth]{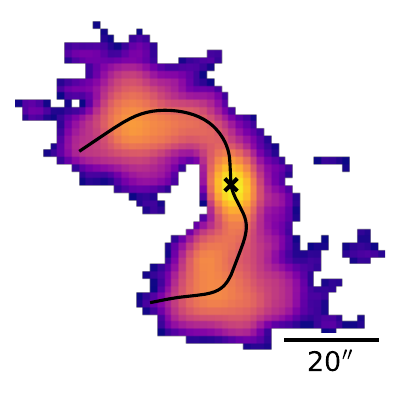} \\

    FRII\\
    \includegraphics[width=0.23\linewidth]{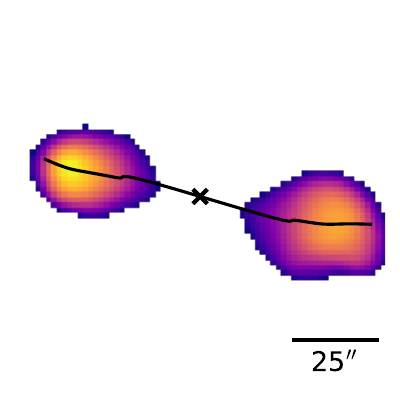}
    \includegraphics[width=0.23\linewidth]{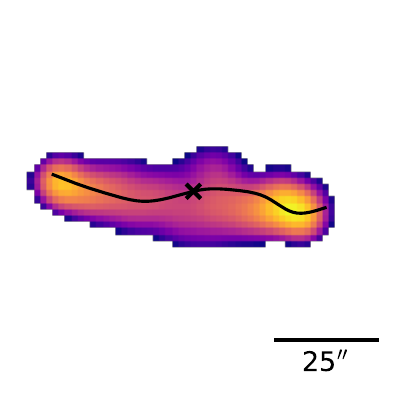}
    \includegraphics[width=0.23\linewidth]{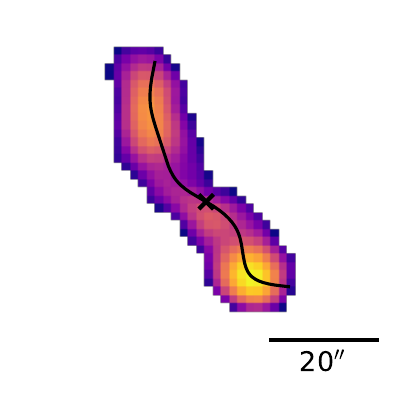}
    \includegraphics[width=0.23\linewidth]{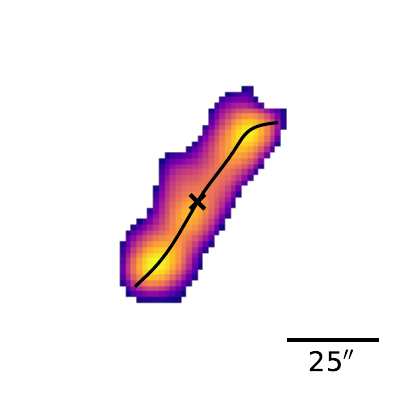} \\

    triple-peaked \\
    \includegraphics[width=0.23\linewidth]{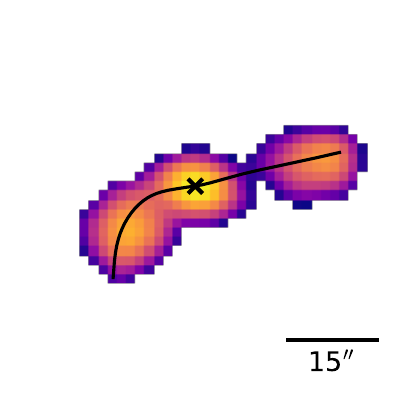}
    \includegraphics[width=0.23\linewidth]{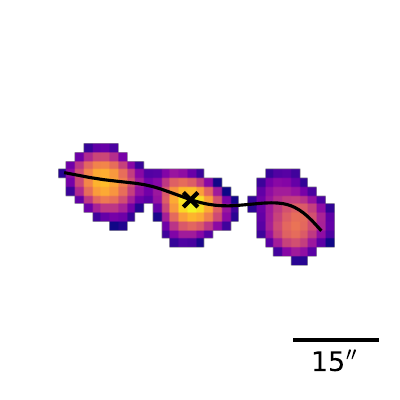}
    \includegraphics[width=0.23\linewidth]{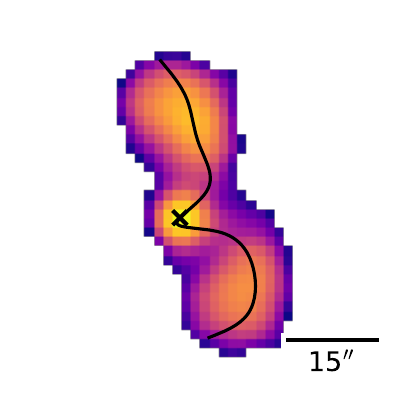}
    \includegraphics[width=0.23\linewidth]{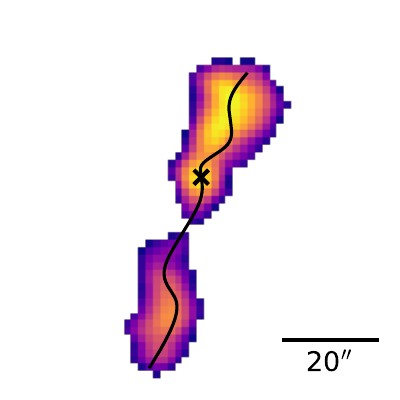} \\

    one-sided ridgelines \\
    \includegraphics[width=0.23\linewidth]{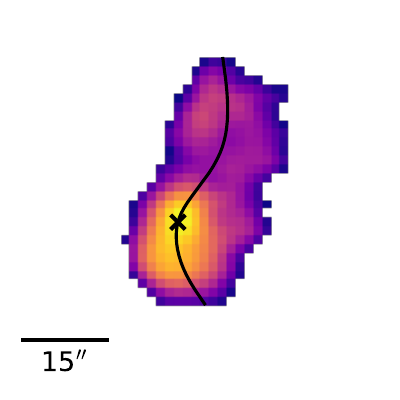}
    \includegraphics[width=0.23\linewidth]{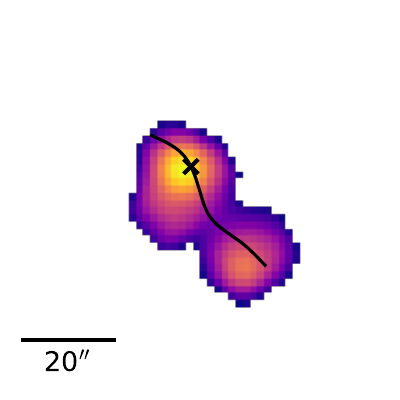}
    \includegraphics[width=0.23\linewidth]{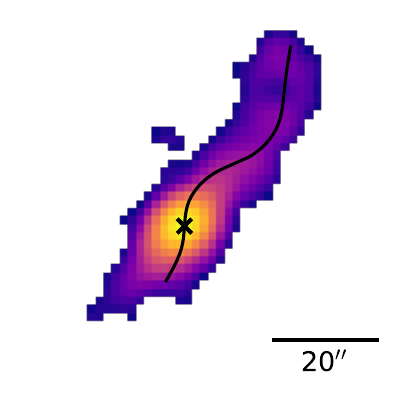}
    \includegraphics[width=0.23\linewidth]{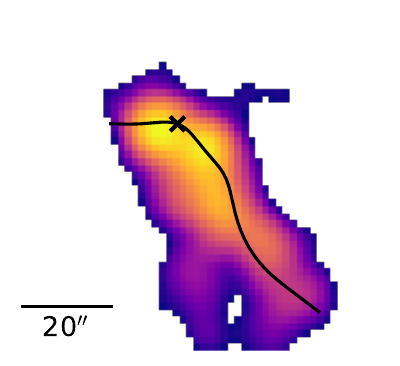} 

    \caption{Randomly selected examples of sources in each morphological class.  The black lines in each figure correspond to that source's ridgeline spline and the black crosses show the closest point on the ridgeline spline to the host galaxy. Each row corresponds to a different class starting with FRIs, then FRIIs, bent, triple and lastly one-sided ridgelines. }
    \label{figure: example images from classes}
\end{figure*}

As our aim is to produce clean groups of sources with similar jet morphologies, through a method that considers the structural and physical properties of the jets, subgroups that were more heterogeneous ($<85 \%$ had the same morphology) were classified as `other'. This class contains $5293$ sources. A random selection of images from each class are shown in Fig. \ref{figure: example images from classes}.

As an additional validation of our method, we compare our catalogue to \cite{2025Jones}, who use visual inspection to classify $2893$ RLAGN from LoTSS DR$2$ with spectroscopic classifications \citep{2024Drake}. Of the FRI and FRII sources in both this catalogue and \cite{2025Jones}, $90\%$ of the classifications are consistent. Of the small proportion of sources that are not consistent, there is no clear class of objects that appears to be missing from this catalogue.

\begin{table*}
    \centering
    \begin{tabular}{l|l|c|c|c|c}
    
        \hline
        Class & Description & Q$1$ & Q$2$ &Q$3$ & Total\\
        \hline
         FRIs & in a centre-brightened SB cluster and host galaxy located within central distribution of GMM. & 2354 & 1032 & 150 & 3536\\  
         
         `bent' FRIs & FRI, $\ge8$ ridge points, $|\kappa_{\rm max}|>0.2$, opposite $\kappa$ direction at ridge ends $\&$ decreasing $\theta_{\rm open}$ profile. & 143 &80 &0 & 223\\ 

         uncertain FRIs & in a centre-brightened SB cluster and host galaxy not within central distribution of GMM.&858&384&53&1295\\ 
         
         FRIIs & in an edge-brightened SB cluster and host galaxy located within central distribution of GMM. & 3590&1598& 222 & 5410\\

         uncertain FRIIs & in an edge-brightened SB cluster and host galaxy not within central distribution of GMM. &786 & 481 &21& 1288\\
        
         triple-peaked & three significant SB peaks &218 &102 &9 &329\\ 

         one-sided ridgelines & one SB peak - contains one-sided sources and ridgelines &1788&942&68& 2798\\

         other & heterogenous subgroups &3491&1622&180&5293 \\
         \hline 
         
    \end{tabular}
    \caption{Final description of each class where Q$1$, Q$2$ and Q$3$ are quality flags corresponding to $\sigma, 2\sigma$ and $3\sigma$ intervals from the mean host location of each surface brightness cluster GMM. Q$1$ sources are considered `good quality' and are used for scientific analysis in Section \ref{results}. Q$2$ and Q$3$ sources are not used in our analysis due to having atypical host locations in comparison to the rest of their surface brightness cluster. }
    \label{Table: description of final classes}
\end{table*}

\subsection{Selection biases}\label{sec: selection effects}
In this section we consider the potential impact of selection effects on our conclusions. We emphasise that the aim of the present work is to examine clean morphological samples, not to compare volume densities for different classes or compute luminosity functions. The complexity of the sample selection means that it is beyond the scope of this work to determine completeness corrections. We have carefully considered the potential impact of the biases summarised below on the science results presented in Section \ref{results}. The main biases in this sample are: (i) radio surface brightness sensitivity, (ii) imposed angular size cut of $45\arcsec$ and (iii) host galaxy selection as a function of redshift.

There are observational selection biases in classifying FRIs and FRIIs with $6\arcsec$ images. FRIIs are easier to resolve due to having two bright hotspots at a distance from the host galaxy. A higher surface brightness sensitivity is needed to identify FRI jets fading out from the bright core \citep{2024Ye, 2024EliasN1}. This can make it difficult to detect fainter FRIs at high redshift and often leads to an underestimation of their physical size. For this reason, and because we cannot account for projection angles, we refer to physical size as observed size. We do not expect the underestimation of FRI sizes to impact our classifications. Changes in source size would alter how the surface profiles are scaled, but the surface brightness of the source  would still decrease roughly monotonically with distance along the ridgeline. We tested the robustness of classifying faint FRIs by artificially lowering the sensitivity limit, step-by-step, until $\sim 97\%$ of the SB profiles fell below the sensitivity limit of the LoTSS images. We found that, unsurprisingly, as sources are scaled to lower surface brightness, the outer parts of FRI plumes drop below the detection limit, changing the LAS and dynamic range of the source. Instead of these being allocated to an incorrect class, $80\%$ of these FRIs would be excluded by the selection criteria (angular size and dynamic range thresholds).  Where sources weren't excluded, they were reassigned to a different FRI-like cluster, with a single-peak mean profile. \cite{2024Jurjen} found similar results in simulating how FRI and FRII morphological classifications change when the same source is observed at different redshifts, altering the observed surface brightness. They found that only $2\%$ of FRIs presented as FRIIs. We note that in addition, the results we present in later sections do not depend on physical source size.

To resolve the smallest FRIs (below $<45 \arcsec$), a balance would need to be found between angular resolution and surface brightness sensitivity \citep{2024Ye,2024EliasN1}. We imposed a $> 45 \arcsec$ cut in order to retain as many sources with well sampled ridgelines for morphological analysis and distinguishable features in their surface brightness profiles for HDBSCAN to identify as possible. We highlight here that this sample is likely missing a proportion of small FRIs. Very long baseline interferometry (VLBI) with LOFAR at various angular resolutions ($0.3 \arcsec, 1\arcsec$ and  $1.2\arcsec$) would be the best way to ensure a more complete sample.

The redshift distributions of H$25$ are shown in Fig. $7$ of \cite{2025Hardcastle}. There is a steep decline in AGN listed in H$24$ above $z \sim 1.2$ as massive galaxies are no longer detected by the Legacy survey above this redshift. Host galaxy identifications become increasingly difficult at high redshift. High luminosity sources above $z>1.5$ have been found to have a higher prevalence of bent morphologies and are less likely to be linear FRIIs \citep{Miley_2008}. However, we do not attempt to draw conclusions about the relative prevalence of different morphologies for any redshift ranges in the present work.

\begin{figure}
	\includegraphics[width=\columnwidth]{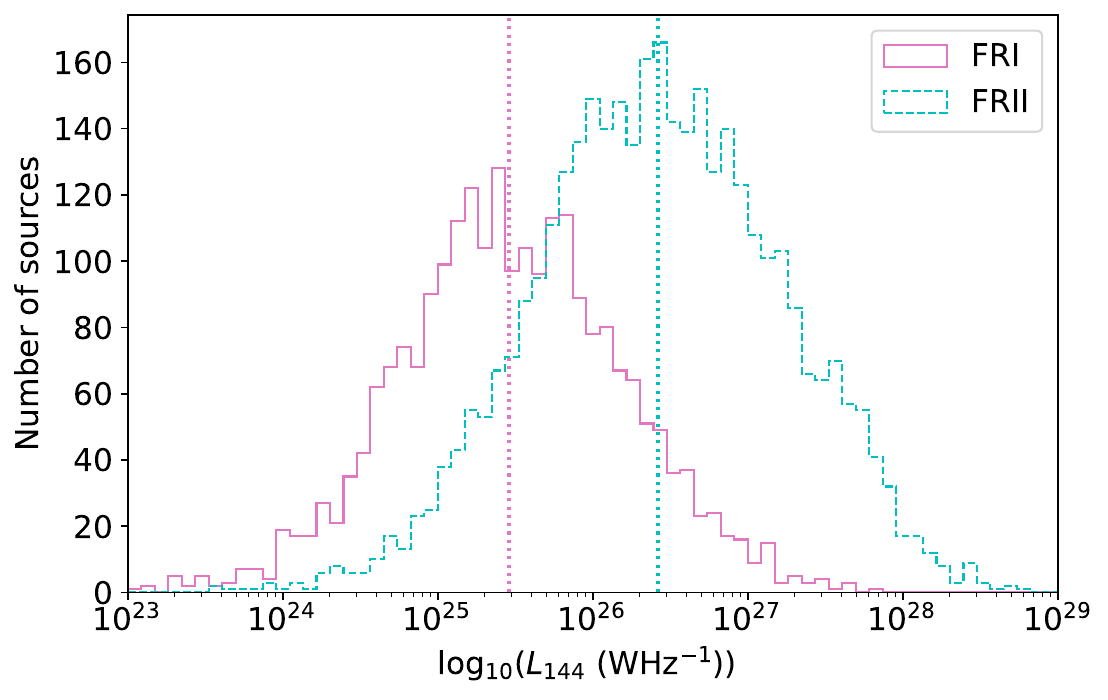}
    \includegraphics[width=\columnwidth]{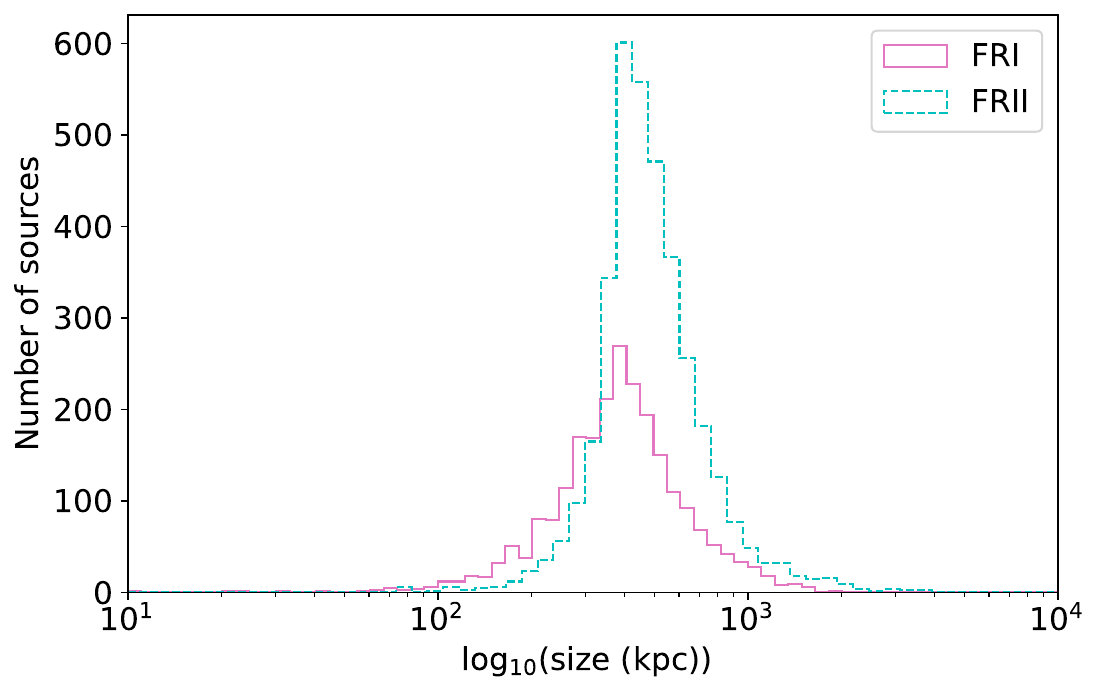}
    \includegraphics[width=\columnwidth]{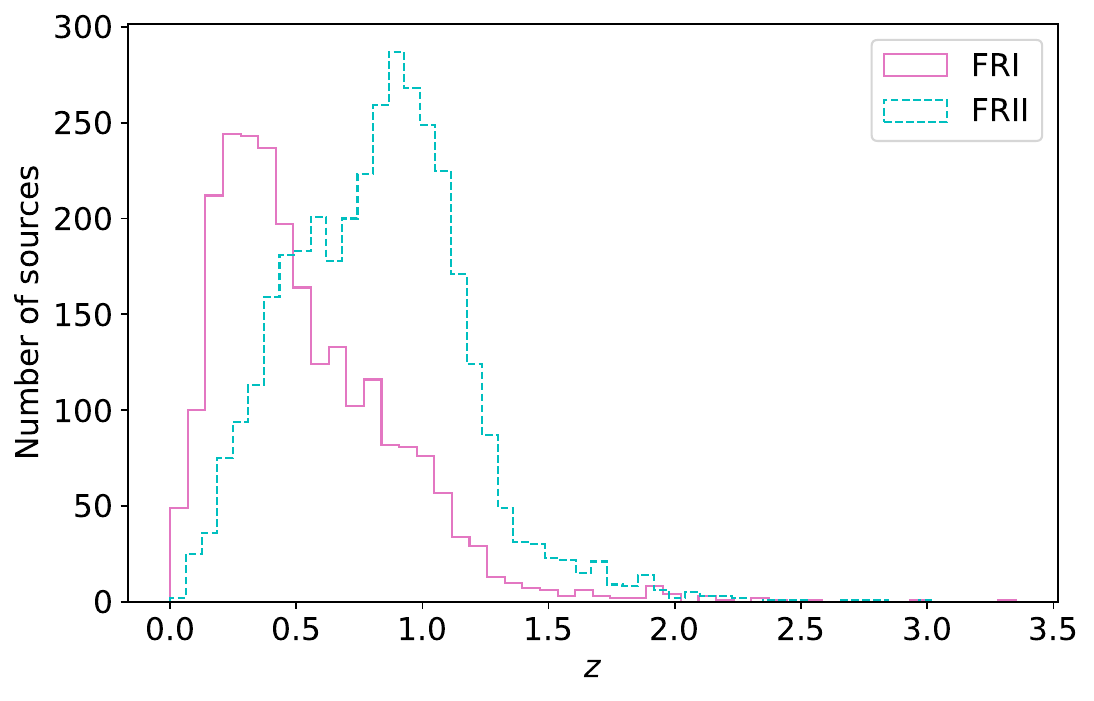}
    \caption{Radio luminosity at $144$ WHz$^{-1}$ (top), observed physical size in kpc (middle) and redshift (bottom) distributions for the Q$1$ FRI (solid pink) and FRII (dashed cyan) morphologies across the full redshift range of this sample. The vertical dotted lines in the top plot represent the median $144$ WHz$^{-1}$ luminosities of the two classifications.}
    \label{fig: FRI+FRII properties}
\end{figure}

Fig. \ref{fig: FRI+FRII properties} (bottom) shows the redshift distribution of sources classified as Q$1$ FRIs and FRIIs by our method. The two distributions are inherently different. The FRI sources are most commonly found at lower redshifts than the FRIIs. Above $z\geq0.8$ the proportion of FRIIs begins to greatly exceed the proportion of FRIs. For these reasons, we will restrict our analysis to sources with $z\leq 0.8$ as described in Section \ref{results}. We consider the impact of selection effects introduced by the different redshift distributions of our morphological classes on our conclusions in later sections.

\section{Results}\label{results}
In this section we report the radio properties of the Q1 (good quality) populations found from our classification method in Section \ref{results: FRI and FRII properties} before investigating their host-galaxy properties in Section \ref{results: host galaxy properties}.

\subsection{FRI and FRII radio properties} \label{results: FRI and FRII properties}

 Fig. \ref{fig: FRI+FRII properties} shows that there is a large amount of overlap between size and luminosity distributions for the FRI and FRII populations. Substantial overlaps between the luminosity distributions of FRIs and FRIIs have only been observed in recent years \citep{Best2009, 2010Gendre,Gendre2013,Capetti2017,Mingo2019, Mingo2022, 2024Jurjen}. Until then, it was thought that nearly all high-luminosity sources were FRIIs and low-luminosity sources were FRIs. The traditional luminosity break is around $L_{144} \sim 10^{26}$ WHz$^{-1}$ \citep{F+R1974}. In our sample the median $144$ MHz luminosities are $2.8 \times 10^{25}$ WHz$^{-1}$ and $2.6 \times 10^{26}$ WHz$^{-1}$  for the FRIs and FRIIs respectively, and the median observed sizes are $384$ and $463$ kpc, respectively. Limiting our sample to $z \leq 0.8$, the median $144$ {W}Hz$^{-1}$ luminosities become  $1.8  \times 10^{25}$ WHz$^{-1}$ and $7.9  \times 10^{25}$ WHz$^{-1}$ respectively, while the median observed sizes become $350$ and $413$ kpc.

 For the full redshift range of our sample, almost a third ($1089$ or $30.3\%$) of the sources in this sample classified as FRIIs have $L_{144} <10^{26}$  WHz$^{-1}$, with $128$ ($3.6\%$) having $L_{144} <10^{25}$  WHz$^{-1}$. When limiting to $z \leq 0.8$, the proportion of low-luminosity FRIIs increases to $56.7 \%$. We will refer to FRIIs with $L_{144} <10^{26}$  WHz$^{-1}$ as `FRII-lows'. 
 
 There also exist a number of Q$1$ FRIs that lie above the traditional luminosity break - $23.3\%$ of FRIs for the full redshift distribution and $9.7 \%$ for $z \leq 0.8$.  A small number of luminous FRIs are known to exist in the $3$CRR and \cite{Mingo2019, Mingo2022} catalogues. We will refer to FRIs with $L_{144} >10^{26}$  WHz$^{-1}$ as `FRI-highs'.

 We now compare the FRI and FRII radio properties of this sample to other catalogues, in particular \cite{Mingo2019}, where the authors use a semi-automated method to classify $5805$ RLAGN from LoTSS DR$1$. It is important to note that the ratios of FRIs to FRIIs in this work ($3:5$) differs significantly to that of \cite{Mingo2019}, where the ratio is $3:1$. For this method, which is based on automatically identifying features in surface brightness profiles deduced by RL-Xid, it is understandable that a higher proportion of FRIIs is identified, given their two distinct hotspots. It is also likely that many of our uncertain or more contaminated classes (e.g. one-sided ridgelines)  contain sources that \cite{Mingo2019} would have classified as FRIs. We emphasise that it was not the aim of this work nor that of \cite{Mingo2019} to estimate volume densities of FRIs or FRIIs, and both morphological samples are incomplete in different ways.
 
 When considering Q$1$ sources at $z \leq 0.8$, the two populations have similar proportions of FRII-lows and FRI-highs. The FRI median luminosity of our sample is very similar to the values found by \cite{Mingo2019} and \cite{2024Jurjen}, despite differences in the relative numbers of sources, method and selection criteria. The median FRII luminosity of our sample is in between the values found by \cite{Mingo2019} and \cite{2024Jurjen}. We also reproduce some of the same conclusions as \cite{Mingo2019} in Section \ref{results}.

\subsection{Host galaxy properties}\label{results: host galaxy properties}

\subsubsection{Radio morphology and stellar mass}\label{sec: radio morph and stellar mass}

Radio morphology cannot be determined by jet power alone, demonstrated by the existence of large numbers of FRII-lows in this work and that of \cite{Mingo2019, Mingo2022}. A fundamental question is then whether differences in radio morphology are instead driven by properties of the host galaxy, large-scale environment or a combination of factors. 

In this section we examine the relationship between radio morphology and stellar mass. We take stellar mass estimates from LoTSS DR$2$ \citep{2023HardcastleLoTSShosts}, which were calculated by adopting the approach of \citet{2022Duncan}. Briefly, stellar masses were estimated using spectral energy distribution (SED) modelling with code previously used by \cite{2014Duncan,2019Duncan, 2021Duncan}. These calculations were subject to certain assumptions about star formation history, metallicity, effects of nebular emission and dust attenuation and are described by \cite{2023HardcastleLoTSShosts}. The completeness of the stellar mass measurements is shown in Fig. $14$ of  \cite{2023HardcastleLoTSShosts}.

\begin{figure*}

    \includegraphics[width=0.47\linewidth]{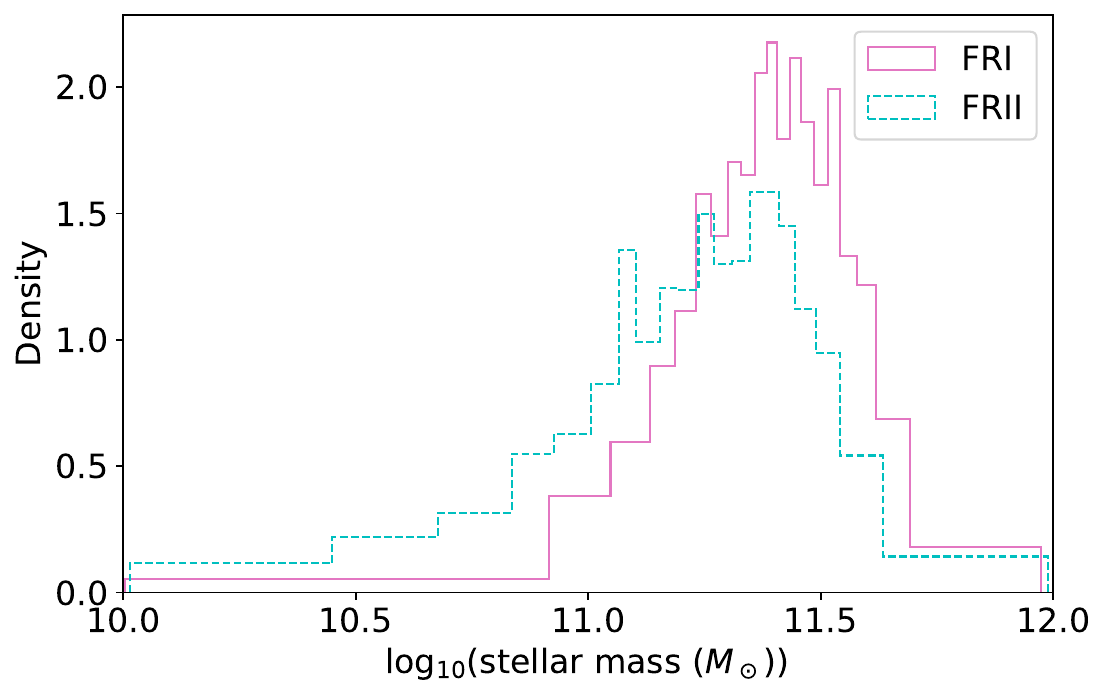}
    \includegraphics[width=0.47\linewidth]{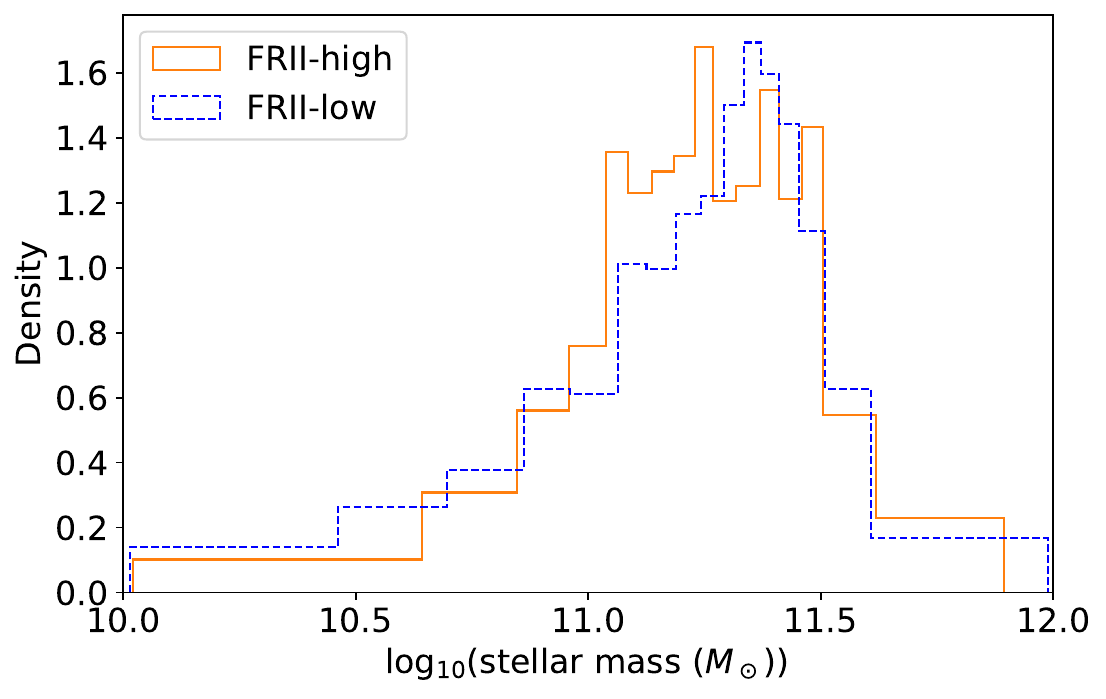}
   \caption{Left: Normalised distribution of host galaxy mass in $M_\odot$ for FRIs (pink) and FRIIs (cyan) with $z \leq 0.8$. Right: The same distribution for FRII-highs (orange) and FRII-lows (blue). }
   \label{fig: mass FRI v FRII and FRII-low v FRII-high}
\end{figure*}

Fig. \ref{fig: mass FRI v FRII and FRII-low v FRII-high} shows the distribution of host galaxy masses for sources classified as FRIs and FRIIs (left) and  FRII-highs and FRII-lows (right) by this work. The majority of sources in this sample have host galaxy masses $10.5 \leq \log_{10}(M/M_\odot) \leq 12$, consistent with the completeness of stellar mass estimates by \cite{2023HardcastleLoTSShosts}. While there is significant overlap between the host galaxy mass distributions of FRIs and FRIIs, differences can be seen in the most and least massive host galaxies. Massive galaxies ($>11.5$) are more likely to host an FRI, whilst the lower mass galaxies ($<11.25$) are more likely to have FRII morphology. There is a tail in the distribution of FRII-lows at low host galaxy mass, similar to that of \cite{Mingo2019}. Including Q$2$ sources in this analysis does not significantly change the overall shapes of these distributions.

{In order to ensure that this result is not caused by the different redshift distributions for FRIs and FRIIs noted in Section \ref{sec: selection effects}, in Fig. \ref{fig: mass FRI v FRII-low} we show the results of selecting a subsample of FRII-lows and FRIs near the FR luminosity break with $10^{25} \le L_{144} \le 10^{26} $ WHz$^{-1}$ and $200 <$ observed size $<1000$ kpc for $z \leq 0.8$ (top) and three different redshift bins (bottom) in our sample. The relationship between host stellar mass and radio morphology then becomes very clear. For intermediate radio luminosities, we find the likelihood of a RLAGN exhibiting FRI or FRII morphology is dependent on host galaxy mass. For masses greater than  $ \sim 11.25$, the probability of forming an FRI becomes greater than the probability of forming an FRII-low. To test whether the stellar mass distributions for the two source populations are significantly different, we performed both a 2-sample Anderson-Darling test \citep{scholz1987} (which is more  sensitive to the tails of the distribution) and a Kolmogorov-Smirnov test \citep{Massey1951TheKT}. For all redshift bins and for the unbinned sample, we  found that the null hypothesis - that FRI and FRII-low subsamples are drawn from the same population - can be rejected at the $> 99\%$ level in both tests. In our sample, the correlation between FR class and host galaxy stellar mass is highly unlikely to be a redshift selection effect. Including Q$2$ sources in this analysis does not change this result.

We do not believe that the observed differences in stellar mass at the FR luminosity boundary can be attributed to environmental boosting of radio luminosity. \cite{2018Hardcastlesim} found that the impact of environment on radio luminosity for jets of the same power is only a factor of $2-3$ over a typical range of environments. In addition, the particle content of the FRIs and FRII-lows may be different \citep[see e.g.][]{croston2018}. Differences in particle content would decrease the radio luminosity of FRIs for a given jet power compared to FRIIs. We examined the group and cluster environments of the sources in our sample in the luminosity range near the FR break used for the analysis presented in Fig. \ref{fig: mass FRI v FRII-low}, using the LoTSS DR2 environmental richness catalogue of \cite{JudeEnvs}. We find no evidence for a difference in the environmental distributions of the Q1 FRI and FRII-lows in the FR break luminosity range, with a Kolmogorov-Smirnov test \citep{Massey1951TheKT} showing that the null hypothesis that the two subsamples have the same parent population cannot be ruled out at $95\%$ confidence level. We therefore conclude that environmental differences between the FRIs and FRII-lows are unlikely to be causing a mismatch in the jet powers being compared.


However, there are a number of FRII-lows in high-mass galaxies, as there were in the sample of \cite{Mingo2022}. This supports the theory that FRII-lows are likely to be a heterogeneous population. RLAGN are known to go through phases of activity (or duty cycles). Lobe morphology is typically determined on timescales of the order of tens to hundreds of Myrs (see e.g. \citealt{2012Konar}, Fig. 14 of \citealt{2018Hardcastlesim, 2019Nandi}). Some FRII-lows in high-mass galaxies could be fading FRII-highs that are no longer active, or whose jets have significantly powered down. Another consideration is the gas density and pressure in the inner kpc region of the host galaxy. If these are low, it could allow a low-power FRII to remain collimated. 

\begin{figure*}
    \centering
    \includegraphics[width=0.5\linewidth]{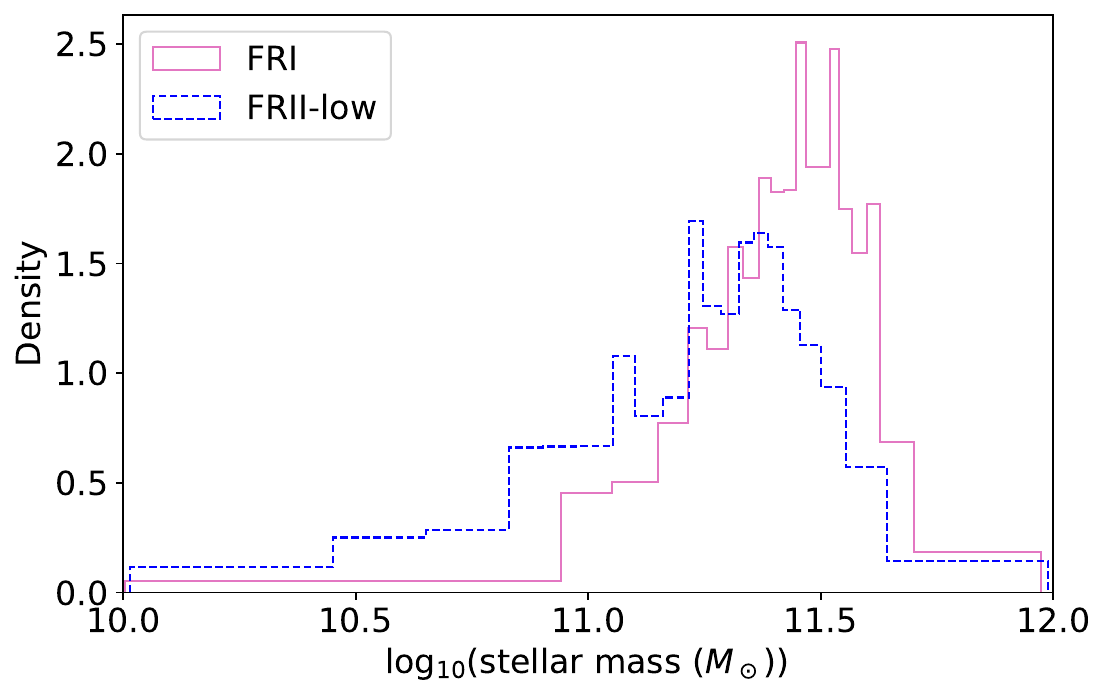}
  \includegraphics[width=\linewidth]{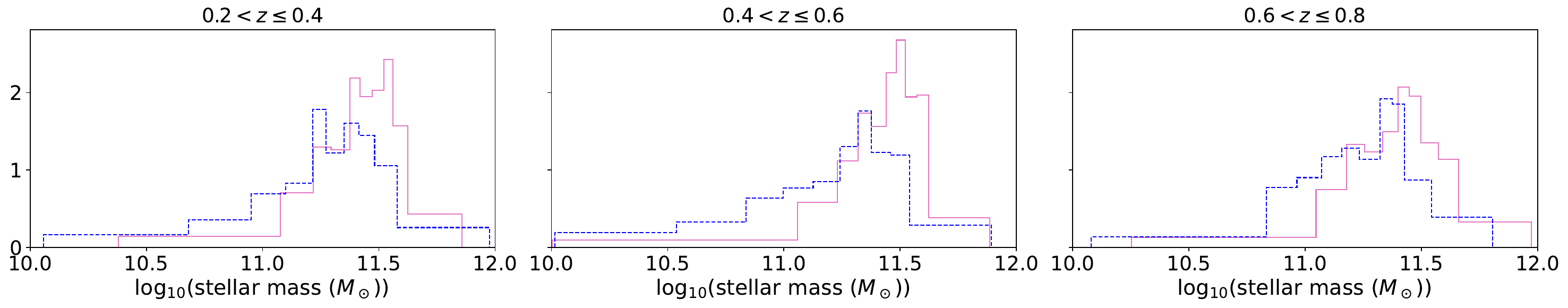}
   \caption{Top: Normalised histogram comparing the stellar mass distribution of FRIs (pink) and FRII-lows (blue) with $10^{25} \le L_{144} \le 10^{26} $ WHz$^{-1}$ and $200 <$ size $<1000$ kpc and $z \leq 0.8$. Bottom: The same plot for three different redshift bins.} 
   \label{fig: mass FRI v FRII-low} 
\end{figure*}

\subsubsection{Exploring FR break dependence}\label{subsection: Host galaxy mass and the FR break}

\cite{Ledlow+Owen1996} were first to suggest that the FR luminosity break was dependent on host galaxy absolute magnitude, used as a proxy for stellar mass \citep[see e.g.][]{2011Konishi}. We now test this theory using the LoTSS DR2 stellar mass estimates. For the Q$1$ FRIs and FRIIs at $z \le0.8$ in our sample (see Fig. \ref{fig: ledlow owen plot (mass)}; top), we observe a slight increase in the FR break luminosity with stellar mass. To test whether this trend could be caused by selection effects,  we examined it for three different redshift bins and performed a partial correlation test, see Fig. \ref{fig: ledlow owen plot (mass)} (bottom). We obtained $r=0.2$ and $p=0.5$, meaning that if redshift were a constant, no noticeable correlation would be found. Including Q$2$ sources in this analysis does not change this result. For this sample, we cannot rule out that selection effects are the cause of the observed trend between the FR break luminosity and stellar mass.

\begin{figure*}
    \centering
    \includegraphics[width=0.5\linewidth]{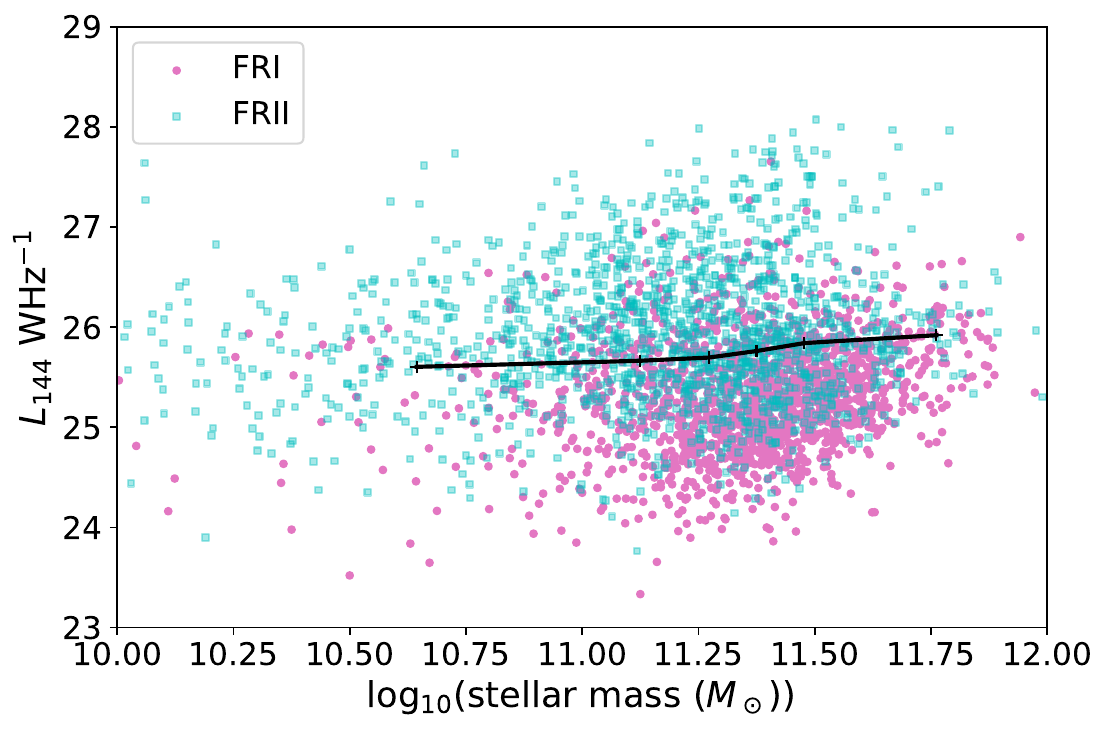}\\
    \includegraphics[width=\linewidth]{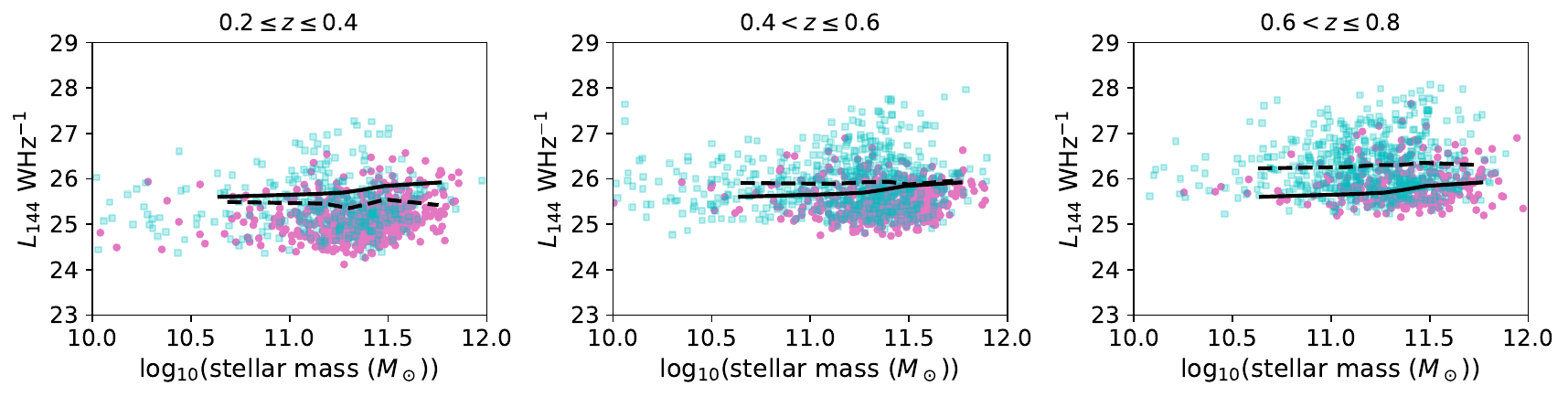}
    \label{fig:ledlow-owen plot}
    \caption{Top: The relationship between radio luminosity and stellar mass for the Q$1$ FRIs and FRIIs in this sample at $z\leq 0.8$. The solid black line indicates the luminosity above which the normalized probability of finding an FRII (cyan  squares) exceeds that of finding an FRI (pink circles) in this sample. This is calculated across $6$ equal-frequency bins between $10.25 \leq M \leq 12.0 M_\odot $. Bottom: The same plot for three different redshift bins, with dashed lines indicating the FR break luminosity for each redshift slice. The solid black line shows the full sample relation as in the upper plot.}
    \label{fig: ledlow owen plot (mass)}
\end{figure*}

 To eliminate the possibility that the observed lack of Ledlow-Owen relation in this sample is caused by systematic or selection effects in stellar mass estimates or by differences in the redshift distributions of the FRI and FRII subsamples (see Section \ref{sec: radio morph and stellar mass} for a description of assumptions made when calculating stellar mass), we consider whether there is a relation between the FR break luminosity and host-galaxy (rest-frame) k-band magnitudes, $K_{s}$, taken from LoTSS DR2 \citep{LoTSSDR22022, 2023HardcastleLoTSShosts}.

For all Q$1$ FRIs and FRIIs in this sample at $z \le 0.8$, a noticeable increase in the FR break luminosity with increasing absolute magnitude is observed, see Fig. \ref{fig: ledlow owen plot (krest)} (top). This is statistically significant according to the Spearman's rank correlation at $>99\%$ confidence level. To determine if this relation was caused by selection effects, we examined it for three different redshift bins and performed a partial correlation test, see Fig. \ref{fig: ledlow owen plot (krest)} (bottom). We found that if redshift were constant, only a weak positive correlation ($r=0.5$) between the FR break luminosity and $K_{s}$ would remain at the $97\%$ significance level. For this sample, we cannot rule out that the observed dependence of the FR break luminosity on $K_s$ is due to redshift selection effects. Including Q$2$ sources in this analysis does not change this result. We discuss our interpretations of this in Section \ref{discussion}.

\begin{figure*}
    \centering
    \includegraphics[width=0.5\linewidth]{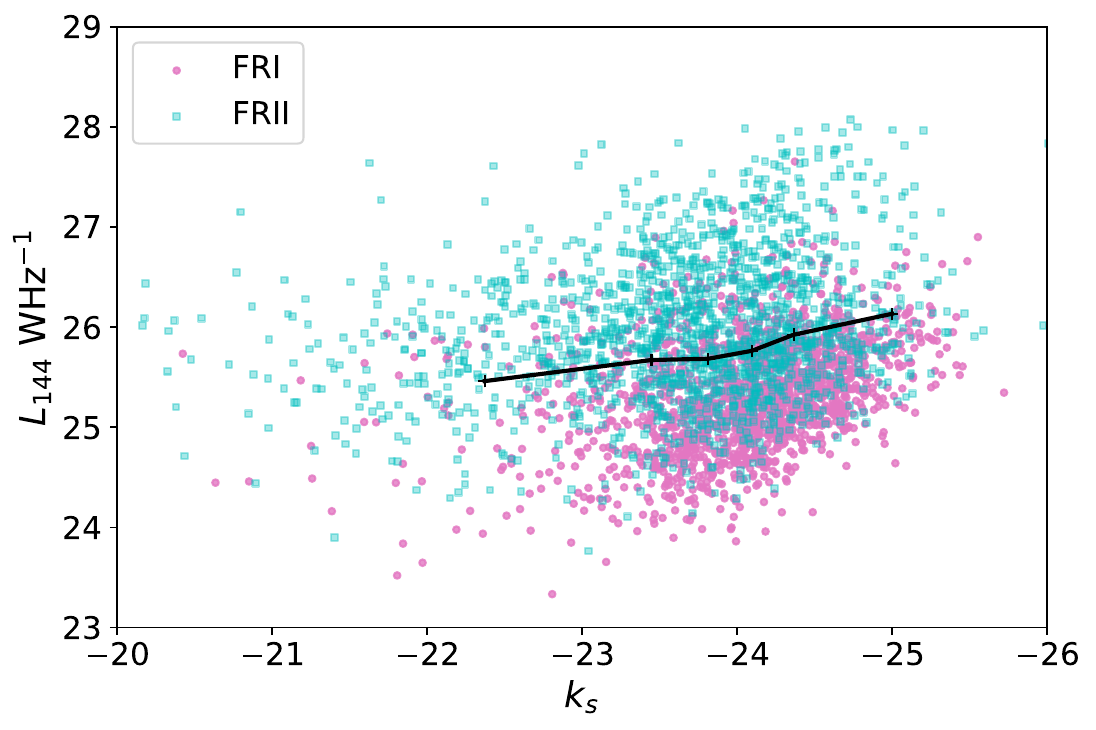}\\
    \includegraphics[width=\linewidth]{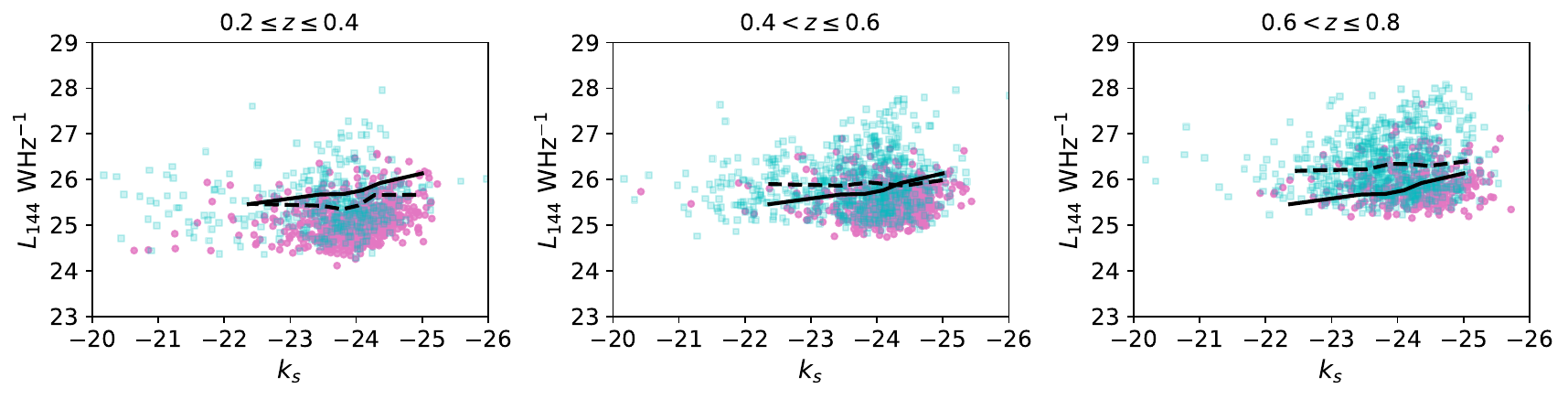}
    \caption{Top: The relationship between radio luminosity and host-galaxy rest frame magnitude, $K_s$, for the Q$1$ FRIs and FRIIs in this sample at $z\leq 0.8$. The solid black line indicates the luminosity above which the normalized probability of finding an FRII (cyan  squares) exceeds that of finding an FRI (pink circles) in this sample.  This is calculated across $6$ equal-frequency bins between $-25.5 \leq K_{rest} \leq -21.5$. Bottom: The same plot for three different redshift bins, with dashed lines indicating the FR break luminosity for each redshift slice. The solid black line shows the full sample relation as in the upper plot.}
    \label{fig: ledlow owen plot (krest)}
\end{figure*}

\subsubsection{Specific star formation rate and radio morphology}\label{section: sSFRs}

FRIs are thought to become mass-loaded due to entrainment and disrupt on kpc scales \citep{Bicknell1994,2002Laing, 2015Wykes, Hardcastle2020}. There are different ways that FRIs can entrain material, one of which is through stellar winds from young stars. Therefore, the type of stellar population within the host galaxy could influence source evolution (see more in Section \ref{introduction}). We now investigate whether small-scale differences in host galaxy properties, such as the type of stellar population, influence radio morphology for sources of similar power. 

We cross-matched the Q$1$ intermediate luminosity ($10^{25} \le L_{144} \le 10^{26} $) FRIs and FRII-lows in this sample, for $z \le 0.8$, with the Max Planck Institute for Astrophysics and Johns Hopkins University optical spectroscopic catalogue \citep[MPA-JHU,][]{2003Kauffmann, 2004Brinchmann, 2004Tremonti, 2007Salim}, based on data from the seventh data release of the Sloan Digital Sky Survey \citep[SDSS DR$7$,][]{2009SDSSDR7}. It should be noted that this subsample is relatively small compared to our whole sample, consisting of $83$ FRIs and $18$ FRII-lows. For this subsample, we take the median specific star formation rate (sSFR) obtained from  within the galaxy fibre aperture using nebular emission lines from MPA-JHU DR$7$ as described by \cite{2004Brinchmann}.  Fig. \ref{Figure:sSFR FRI v FRII-low} shows the distribution of this subsample of FRIs and FRII-lows for different median sSFRs. It then becomes clear that in this subsample, the probability of producing an FRI exceeds the probability of producing an FRII-low in host-galaxies with a higher median sSFR. To test whether the sSFR distributions for the two source populations are significantly different, we performed a 2-sample Anderson-Darling test \citep{scholz1987}. We found that the null hypothesis - that FRII-low and FRI samples are drawn from the same stellar population - can be rejected at the $97\%$ level.

\begin{figure}
    \centering
    \includegraphics[width=\columnwidth]{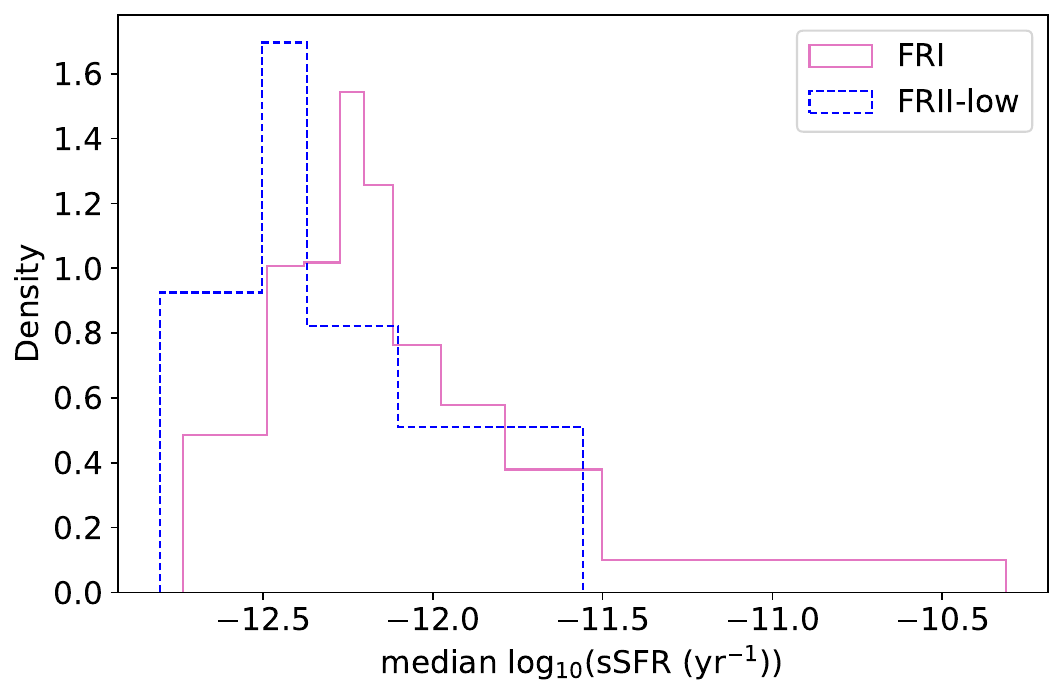}
    \caption{Normalised histograms comparing the median sSFR  distributions, obtained from MPA-JHU DR$7$, of FRIs (pink) and FRII-lows (blue) with $10^{25} \le L_{144} \le 10^{26} $ WHz$^{-1}$ and $z \le 0.8$ in this sample.  }
    \label{Figure:sSFR FRI v FRII-low}
\end{figure}

\section{Discussion}\label{discussion}
In this paper we have described a new automated method to classify RLAGN using ridgelines \citep{Barkus2022} and its application to the LoTSS DR2 AGN catalogue \citep{2025Hardcastle}. This method has identified $2354$ and $3590$ Q$1$ FRIs and FRIIs respectively. At each stage of the classification process we visually inspected $100$ randomly selected sources from each class and report that $ \sim 90\%$ of Q$1$ sources are correctly classified. We compared our FR classifications to those of \cite{Mingo2019} to check their quality. Restricting to $z \leq 0.8$, we find the proportion of FRII-lows and FRI-highs in our sample to be consistent with \cite{Mingo2019}. In addition, the median $144$ MHz luminosities of the FRIs and FRIIs are similar. We are confident that the Q$1$ FRI and FRII classes of sources can be used for science. We also identify `bent', triple-peaked and one-sided sources with this method. We believe that certain improvements to the ridgeline code are needed before using these categories for scientific analysis, as described in Section \ref{final classification}.

We first explored the physical properties of the FRI and FRII sources in this sample. We restricted our analysis to Q$1$ sources with $z \leq 0.8$ based on the redshift distributions of our catalogue and that of the parent sample H$24$. Although the FRIIs do have a higher median luminosity compared to the FRIs, we found a large amount of overlap between their luminosity distributions, supporting previous work that radio luminosity alone is not enough to classify morphology \citep[as observed in other samples][]{Best2009, Gendre2013, Mingo2019}. The FRIIs also have a higher median size than the FRIs, but again, there is a large amount of overlap in the FRI and FRII size distributions. Furthermore, we find that $56.7 \%$ of the FRIIs have $L_{144} < 1 \times 10^{26}$ WHz$^{-1}$. These results suggest that jet power alone is not enough to explain the FR dichotomy.

We instead considered whether morphological differences could be driven by properties of the host galaxy, large-scale environment or a combination of factors. FRII jets are thought to remain relativistic throughout until terminating in a hotspot, while FRI jets are known to be initially relativistic and then become mass-loaded due to entrainment, causing them decelerate on kpc scales \citep{Bicknell1994, 2002Laing, 2015Wykes,Tchekhovskoy2016, Hardcastle2020}. This suggests a relationship between jet power and environment. \cite{Ledlow+Owen1996} found the FR break luminosity to be dependent on host-galaxy magnitude for the $3$C sample \citep{1971Mackay}. This result was initially thought to provide evidence to support a jet power-environment  relationship as the cause of the FR divide. However, it has since been questioned due to severe selection effects \citep{Hardcastle2020}.

We re-examined the Ledlow-Owen relation with our classified catalogue and found no evidence to support it, when considering selection effects. We do not believe this disproves the idea that jets are disrupting due to environmental differences. The original relation was observed on a sample with severe selection biases. It was not known if it held across the full RLAGN population. \cite{Mingo2019} also could not rule out that the observed correlation between the FR break luminosity and host-galaxy absolute magnitude in their sample was due to selection effects. Furthermore, it is now known that the relationship between radio luminosity and jet power is subject to a lot of scatter due to external pressure, radiative losses and differences in particle content of the two morphological types \citep{croston2018, Hardcastle2020}. This means that using radio luminosity as a proxy for jet power when exploring the jet disruption model may not be reliable.

Though the FR break luminosity may not be directly linked to host-galaxy stellar mass or absolute magnitude for this sample, we do find that for sources of similar size ($200 \le$ size $ \leq 1000$ kpc) and luminosity near the FR break ($10^{25} \le L_{144} \le 10^{26} $ WHz$^{-1}$) the probability of forming an FRI exceeds that of forming an FRII-low in massive host galaxies, both for the full sample and for subsamples in three different redshift bins. The host galaxy stellar mass distributions for FRIs and FRII-lows are intrinsically different and this is demonstrated with both a 2-sample Anderson-Darling and Kolmogorov-Smirnov test  at $>99\%$ confidence. This result suggests that inner environment is likely to have a role in determining when jets of a similar power disrupt, consistent with the conclusions of \cite{Bicknell1994, 2002Laing}. 

As stellar populations have been postulated to play a role in jet disruption (see Section \ref{introduction}), we also compared the specific star formation rates of the FRIs and FRII-lows, as presented in Fig. \ref{Figure:sSFR FRI v FRII-low}. We find that the median sSFR of FRIs in this subsample is $1.7 ^{+0.4}_{-0.3}$ times larger than that of the FRII-lows. Though this result is tentative,  we consider here the potential physical implication of our results on the ISM versus stellar wind entrainment models.  First we consider the results of \cite{2015Wykes}, the only example of internal entrainment modelling on observational data. In Centaurus A, mass loading by stellar winds is sufficient to cause deceleration, but by a slim margin. The SFRs we find in this work are a factor of $10$ smaller than that of Centaurus A, so if stellar winds are only just sufficient to cause deceleration in Centaurus A, it is unlikely that this alone is enough to cause deceleration in all of our sources. In addition, we are considering more powerful FRIs near the FR break, for which \cite{2014Perucho} found mass loading by stellar winds to be insufficient. We therefore find it unlikely that mass loading from stellar winds is the cause of jet disruption in this sample. \\ An alternative scenario in which stellar activity could influence jet disruption was presented by \cite{2020Perucho}, who showed that the entry and exit of stars from jets can plausibly create a jet-ISM mixing layer. Basic consideration of a steady-state star formation scenario indicates that the observed sSFR difference between the FRIs and FRII-lows around the FR break luminosity would correspond to a small difference in the proportion of massive stars most relevant for this scenario. More detailed modelling work would be required to determine whether the relatively small difference in typical star formation rates we observe for FRIs and FRII-lows around the FR break could be sufficient to influence the probability of jet disruption.

\section{Conclusions}\label{conclusions}
In this work we have described a new automated method of classifying RLAGN using ridgelines \citep{Barkus2022} and examined the physical and host galaxy properties of the FRI and FRII jets identified by it. Our sample is based on data from the LOFAR Two-Metre Sky Survey's second data release (LoTSS DR2) AGN catalogue \citep{2025Hardcastle} but also uses spectroscopic information from SDSS DR$7$ \citep{2003Kauffmann, 2004Brinchmann, 2004Tremonti, 2007Salim, 2009SDSSDR7}.

Our results have led to the following conclusions:
\begin{enumerate}
    \item Sources with FRI and FRII morphology can be found across a wide range of radio luminosities. On average, FRII sources do have a higher radio luminosity than FRIs, but there is a large amount of overlap.
    \item When considering selection effects, we find no evidence that this sample supports the Ledlow-Owen relation in terms of stellar mass or host-galaxy rest-frame absolute magnitude.
    \item For sources of intermediate observed size and luminosity near the FR break ($10^{25} \le L_{144} \le 10^{26} $ WHz$^{-1}$, $200 <$ observed size $<1000$ kpc), the probability of forming an FRI or FRII is dependent on stellar mass. There are tentative indications that sSFR may also be relevant.
    
\end{enumerate}

The cause of different morphologies in this RLAGN population is complex, but careful consideration of sources near the FR break ($10^{25} \le L_{144} \le 10^{26} $ WHz$^{-1}$) supports the role of inner environment in determining when jets of similar power disrupt. This catalogue is made publicly available and has the potential to enable further exploration into the impacts of jets on their environment. Future plans involve extending this classification method to include subclasses (e.g. lobed FRIs, tailed FRIs, restarting sources), by improving the ridgeline drawing.

\section*{Acknowledgements}

LC acknowledges support from the UK Research \& Innovation (UKRI) Science and Technology Facilities Council (STFC) for a  studentship [grant number ST/Y509449/1]. JHC and BM gratefully acknowledge the support of UKRI STFC under grants [ST/T000295/1] and [ST/X001164/1]. BM acknowledges support from UKRI STFC for an Ernest Rutherford Fellowship [grant number ST/Z510257/1]. BB acknowledges support from STFC for a studentship and under grant [ST/X002543/1]. MJH thanks STFC for support [ST/Y001249/1]. JMGHJdJ recognizes support from project CORTEX (NWA.1160.18.316) of research programme NWA-ORC which is (partly) financed by the Dutch Research Council (NWO). We thank Yifei Gong and Judith Ineson for useful discussions.

LOFAR is the Low Frequency Array designed and constructed by ASTRON. It has observing, data processing, and data storage facilities in several countries, which are owned by various parties (each with their own funding sources), and which are collectively operated by the ILT foundation under a joint scientific policy. The ILT resources have benefited from the following recent major funding sources: CNRS-INSU, Observatoire de Paris and Université d'Orléans, France; BMBF, MIWF-NRW, MPG, Germany; Science Foundation Ireland (SFI), Department of Business, Enterprise and Innovation (DBEI), Ireland; NWO, The Netherlands; The Science and Technology Facilities Council, UK; Ministry of Science and Higher Education, Poland; The Istituto Nazionale di Astrofisica (INAF), Italy.

This research made use of the Dutch national e-infrastructure with support of the SURF Cooperative (e-infra 180169) and the LOFAR e-infra group. The Jülich LOFAR Long Term Archive and the German LOFAR network are both coordinated and operated by the Jülich Supercomputing Centre (JSC), and computing resources on the supercomputer JUWELS at JSC were provided by the Gauss Centre for Supercomputing e.V. (grant CHTB00) through the John von Neumann Institute for Computing (NIC).

This research made use of the University of Hertfordshire high-performance computing facility and the LOFAR-UK computing facility located at the University of Hertfordshire and supported by STFC [ST/P000096/1].
\section*{Data Availability}
The RLAGN classified catalogue resulting from this work will be made available at \url{https://lofar-surveys.org/} on publication.


\bibliographystyle{mnras}
\bibliography{bib} 




\appendix

\section{Host subgrouping}

Sources with the same SB profiles but different host galaxy positions do not belong in the same class, as the location of the host galaxy provides important information
on the formation history and symmetry of the jets (the distance of
each jet from the host galaxy). We examined the distributions of host galaxy positions for each SB cluster and noticed that there were  $1-3$ peaks in the distributions. To separate the common host galaxy locations within the SB clusters,  we used Gaussian Mixture Modelling (GMM). Each Gaussian distribution that the GMM model identified was treated as a subgroup of its SB cluster i.e. for each SB cluster, there were $1-3$ subgroups of host positions, see \ref{fig: host_subgroups} for examples of this.

\begin{figure*}
    \centering
    \includegraphics[width=0.49\linewidth]{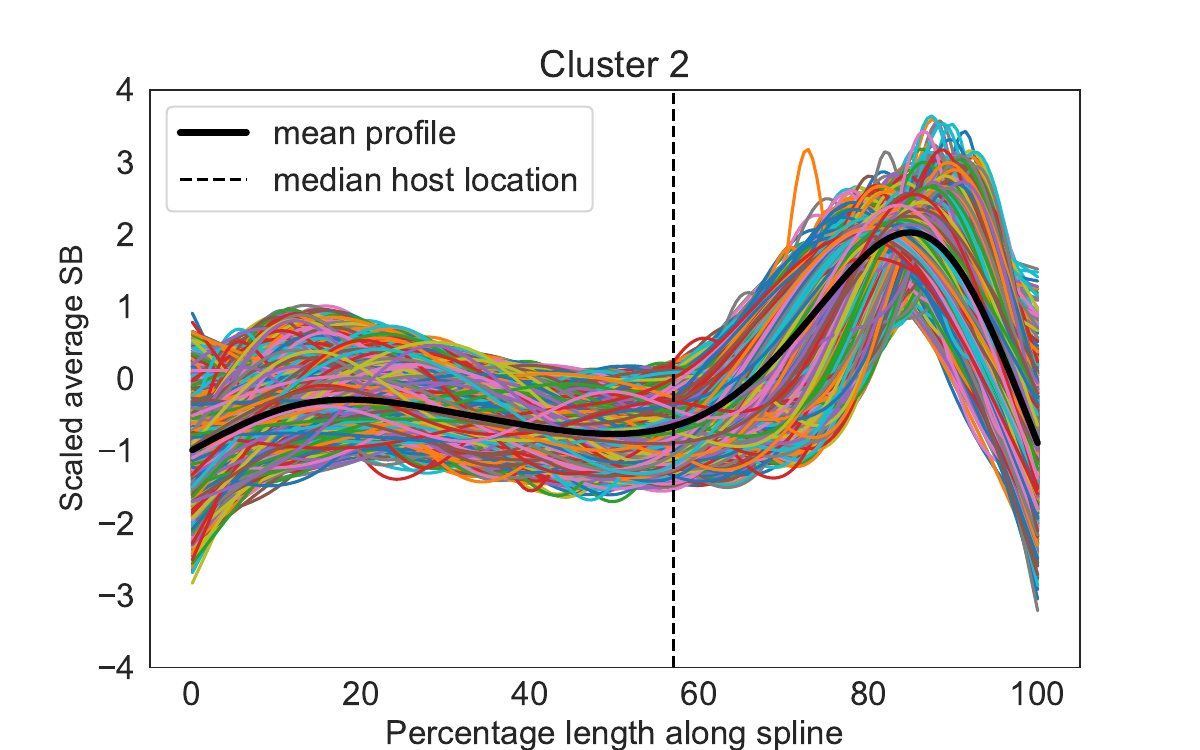}
    \includegraphics[width=0.49\linewidth]{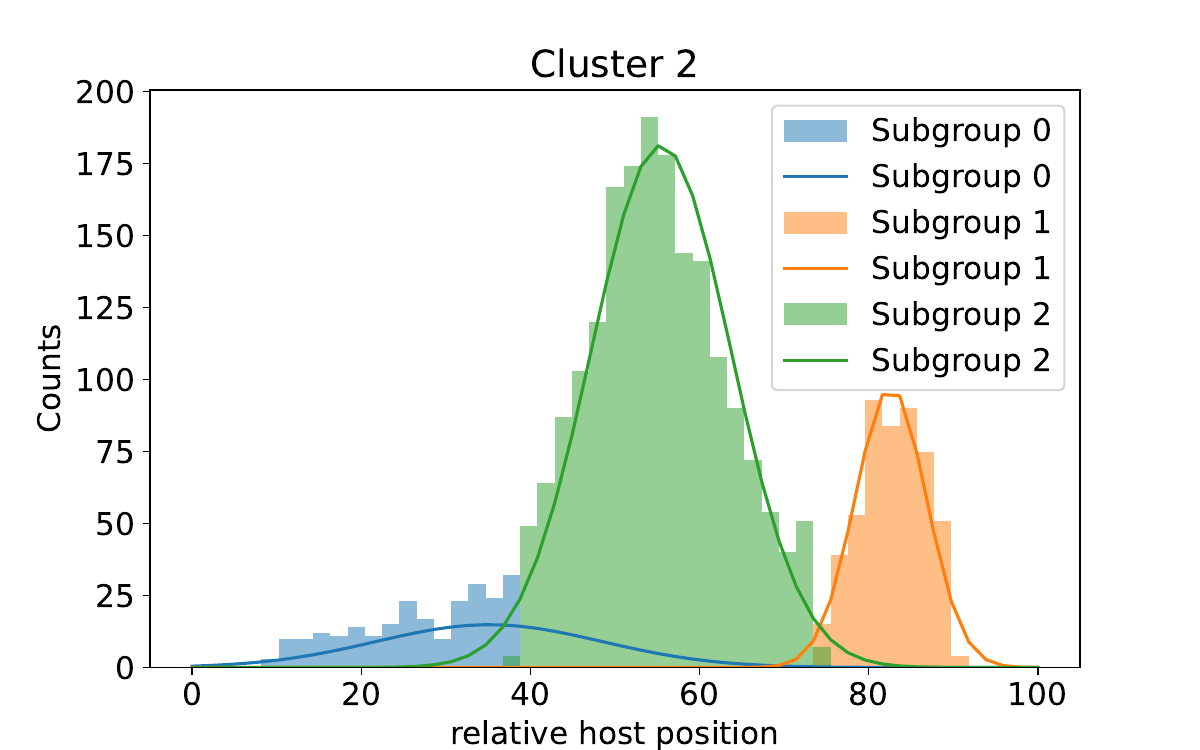}
    \\
    \includegraphics[width=0.49\linewidth]{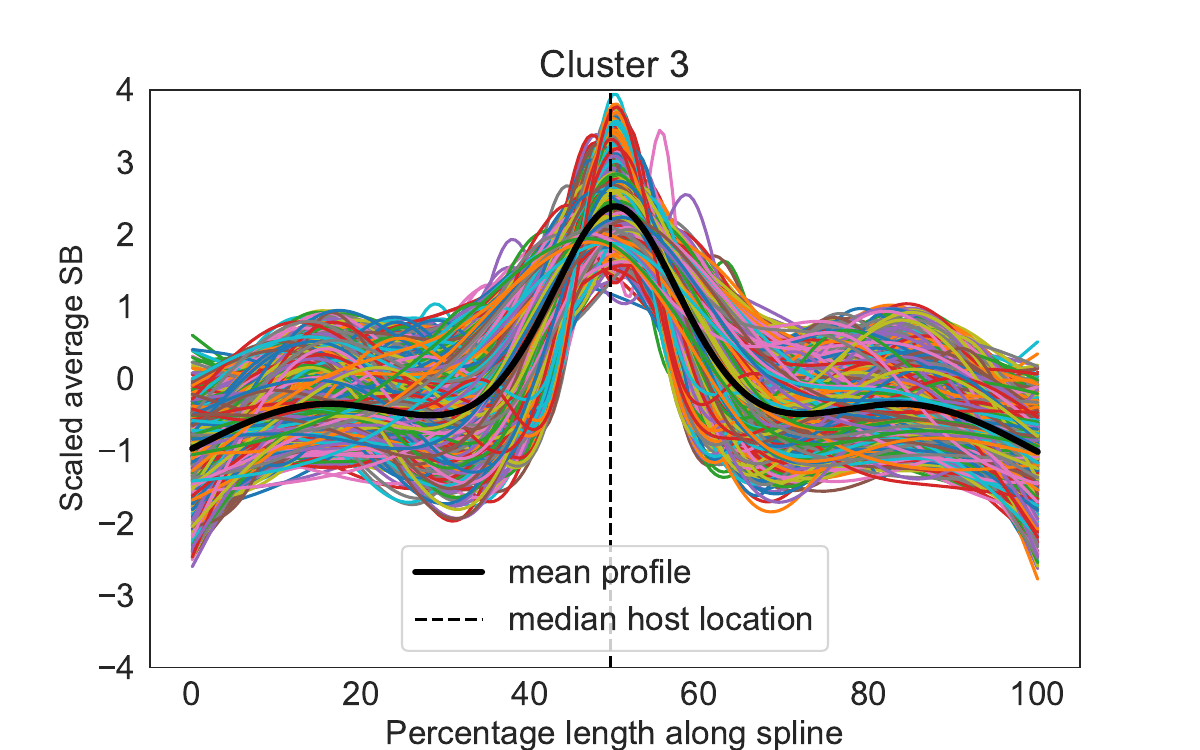}
    \includegraphics[width=0.49\linewidth]{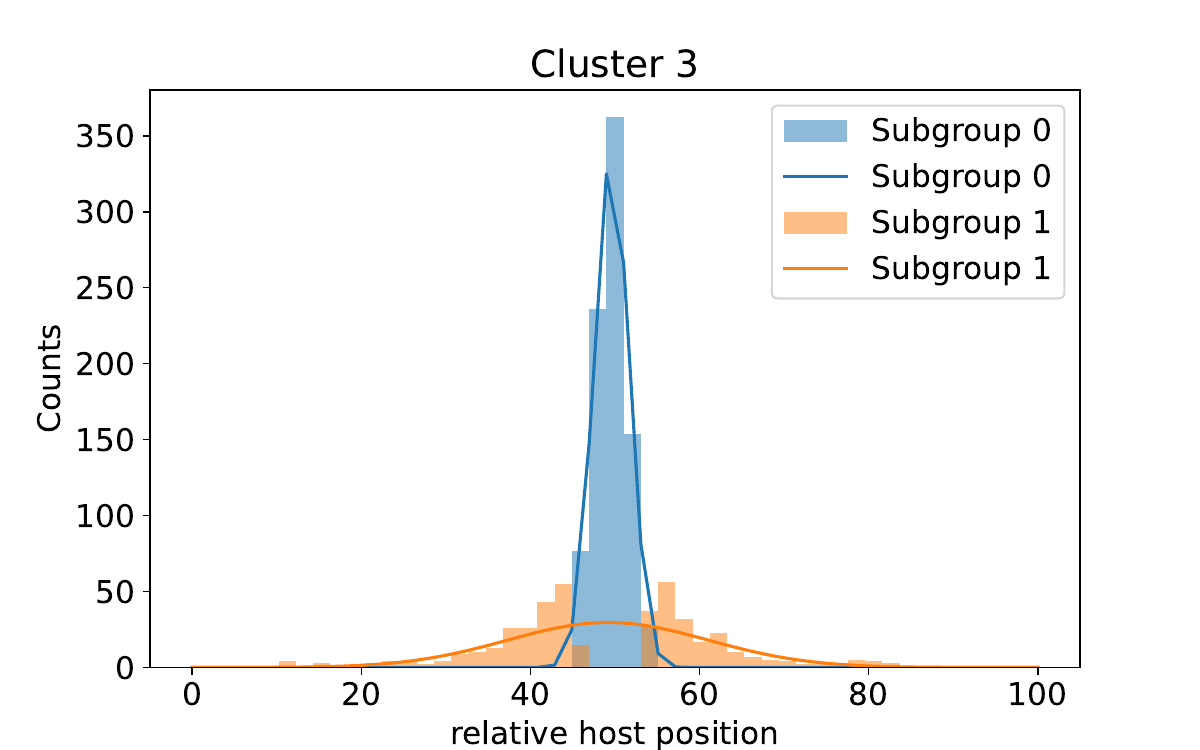}
    \\
    \includegraphics[width=0.49\linewidth]{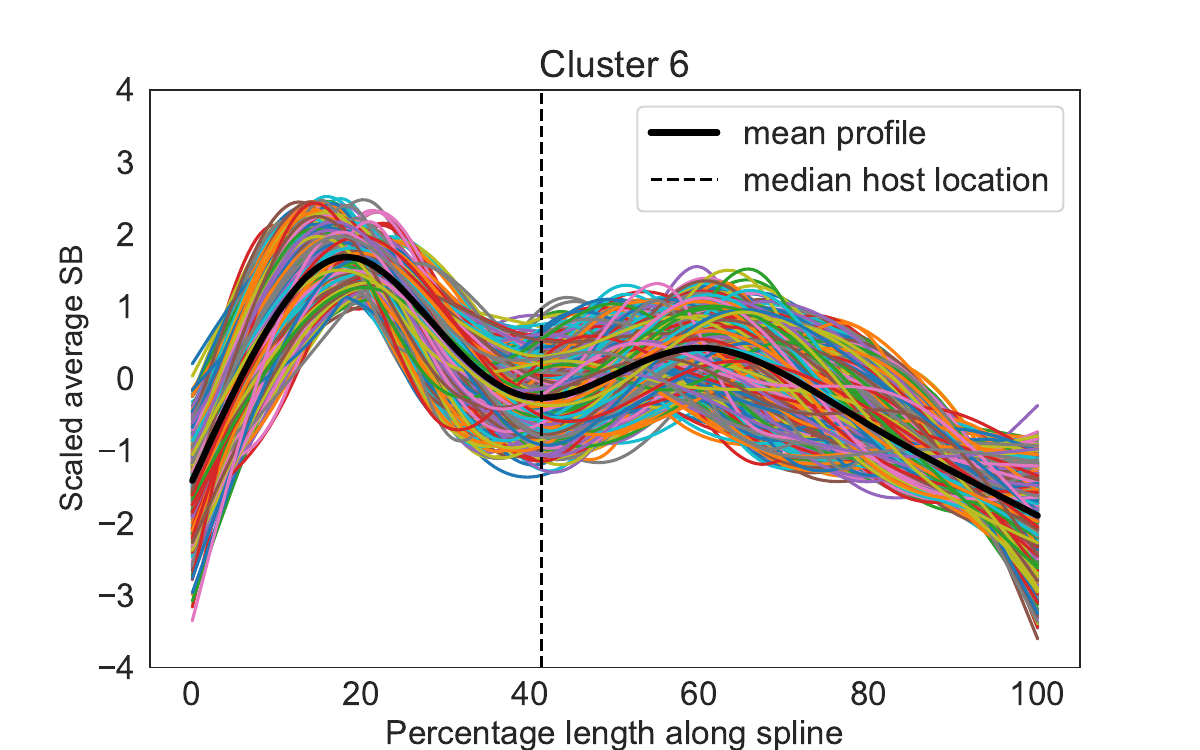}
    \includegraphics[width=0.49\linewidth]{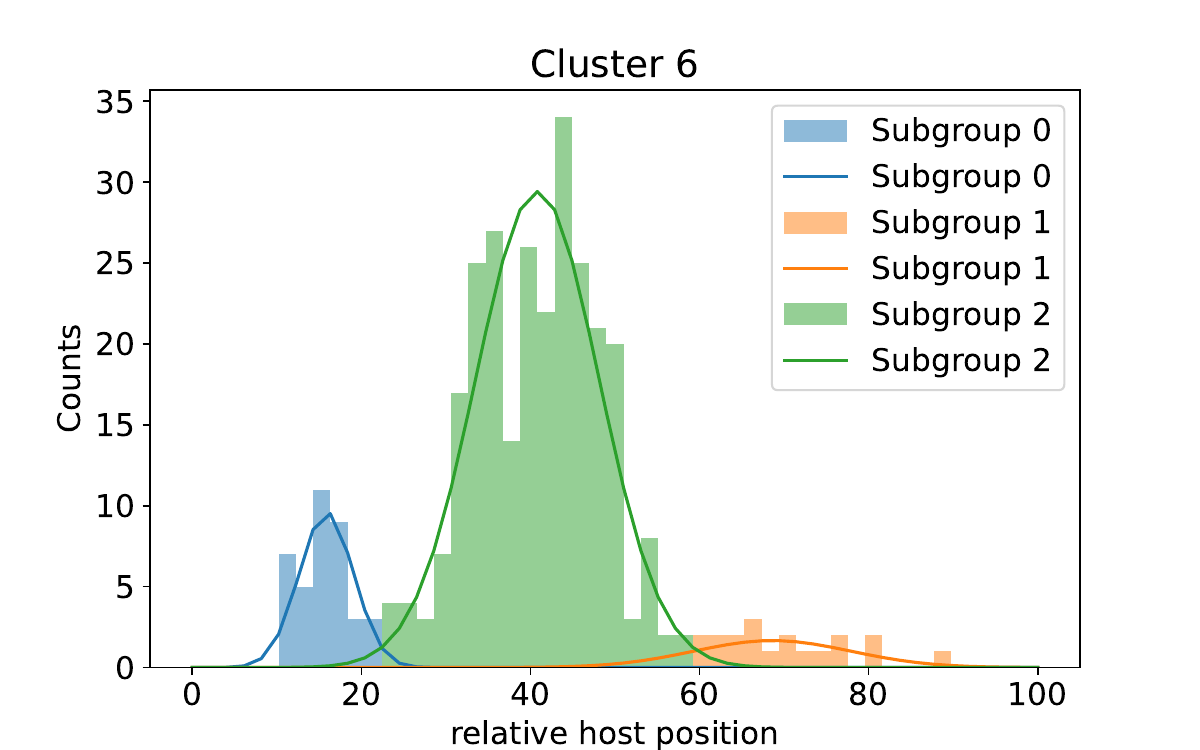}
    \\
    \includegraphics[width=0.49\linewidth]{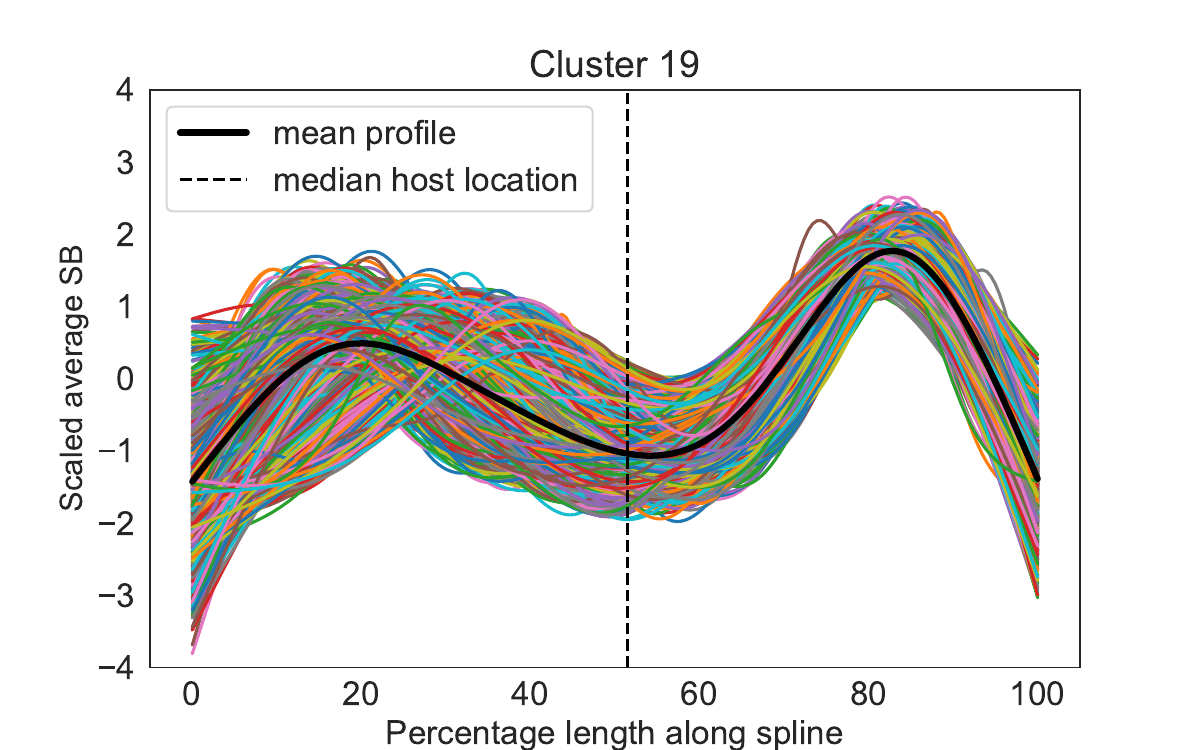}
    \includegraphics[width=0.49\linewidth]{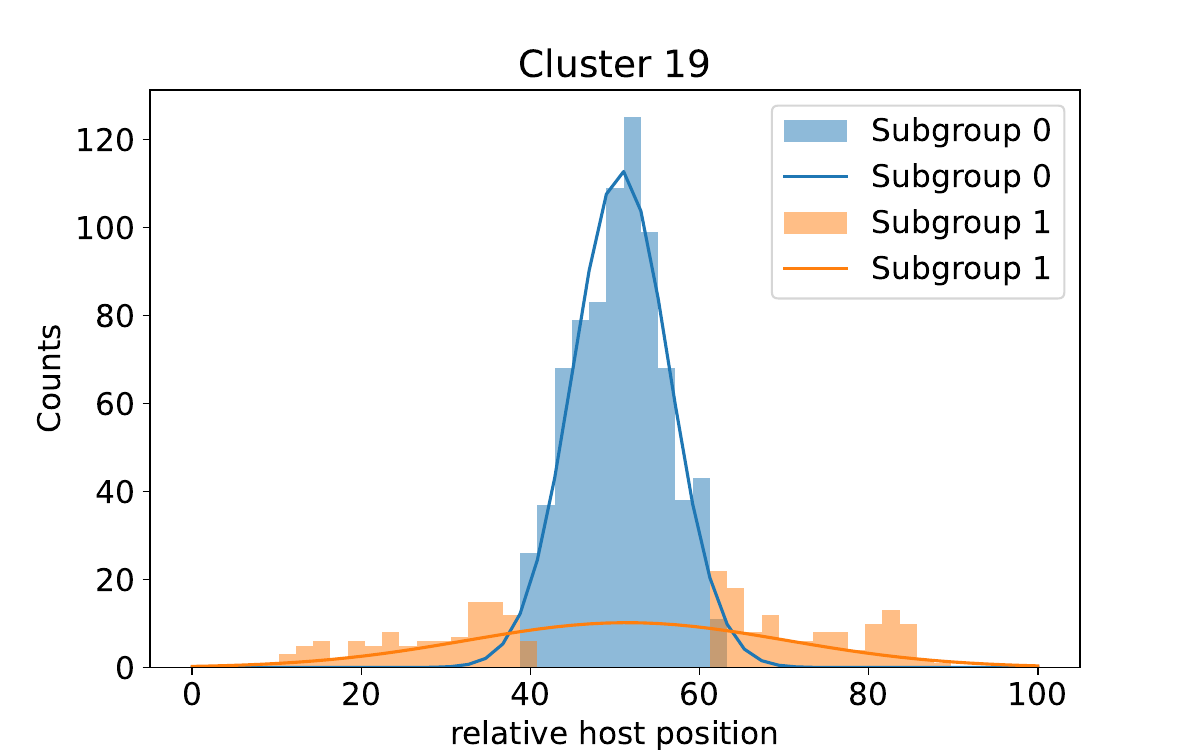}
    \caption{Left: Surface brightness clusters from HDBSCAN. Right: Corresponding host distributions, colour coded by the subgroup number.}
    \label{fig: host_subgroups}
\end{figure*}


\label{lastpage}
\end{document}